\documentclass[5p,onecolumn, natbib,compress]{elsarticle}

\usepackage{amssymb}
\usepackage{amsmath}
 \usepackage{amsthm,amssymb}
\usepackage{caption}  
\usepackage{flushend}
\usepackage{graphicx}                
\usepackage{epstopdf} 
\usepackage{hyperref}
\usepackage{xcolor} 
\usepackage{booktabs}
\usepackage[utf8]{inputenc}  
\usepackage{enumitem}  
\usepackage{stfloats}

\usepackage{array} 

\newcolumntype{C}[1]{>{\centering\arraybackslash}p{#1}}
\newcolumntype{C}[1]{>{\centering\arraybackslash}p{#1}}

\newtheorem{proposition}{Proposition}
\newtheorem{remark}{Remark}

\journal{Digital Signal Processing}

\begin{document}

\begin{frontmatter}



\title{Reconfigurable Intelligent Surface-Enabled Channel Signature Modulation}


%

\author{M. A. Teeti}
\ead{teeti.moh@gmail.com}
\begin{abstract}

This work proposes \emph{RIS-enabled channel signature modulation} (RIS-CSM), a lightweight index modulation scheme for reconfigurable intelligent surfaces (RIS). An $N$-element RIS is partitioned into disjoint groups, each employing predetermined binary reflection patterns to generate distinct channel signatures at an $n_R$-antenna receiver, without RIS-side beamforming. Information is embedded in the indices of these signatures, enabling simple channel estimation and scalable spectral efficiency. A closed-form upper bound on error probability and capacity analysis are derived, revealing diversity order $n_R$ and coding gain proportional to $N$. Simulation results under Rayleigh fading validate the theoretical analysis. Moreover, simulations indicate that spatial correlation among RIS elements can improve system performance at low spectral efficiency.

\end{abstract}

%
%


\end{frontmatter}



\section{Introduction}
\label{sec1}

The growing need for enhanced spectral efficiency (SE) and energy efficiency (EE) in future wireless networks has driven research into innovative technologies such as reconfigurable intelligent surfaces (RIS) and index modulation (IM)~\cite{1580755,8315127,7509396}. RIS, a metasurface with software-programmable elements, dynamically manipulates the wireless propagation environment by smartly altering the characteristics of incoming electromagnetic waves (e.g., phase and amplitude), thereby enhancing received signal power without additional power consumption~\cite{9698029}.

Index modulation (IM) schemes significantly enhance SE and EE by encoding information bits into the indices of communication resources, such as one-dimensional entities including antennas (spatial domain) \cite{wen2019survey}, subcarriers (frequency domain)~\cite{TEETI2023103834}, time slots (time domain), or spreading codes (code domain), as well as combinations of these resources (multi-dimensional entities). This approach enables efficient data transmission without requiring additional bandwidth or power, making IM a compelling solution for modern wireless systems~\cite{basar2017index}. For a detailed exploration of IM techniques and their applications, readers are referred to comprehensive surveys in~\cite{wen2019survey, mao2018survey, 10680066}.

The integration of RIS with IM has emerged as a promising paradigm for future communication systems that combines the benefits of both technologies~\cite{canbilen2020reconfigurable,Yuan2021,rao2024performance}. 
In the literature, the role of RIS can be solely as a beamformer/phase aligner, reflecting impinging signals with adjustable coefficients to optimize the received signal power. Alternatively, RIS can be utilized as an information encoder, with or without phase alignment. When phase alignment is employed at the RIS, it is typically assumed that channel state information (CSI) is available at the RIS.

\textit{RIS as beamformer/phase aligner:}  
In this category, the RIS in RIS-assisted IM schemes acts as a passive, active, or hybrid beamformer \cite{9698029, Yuan2021, 8981888,10214326,Marin2024,10388479,10416245,10685086,10844041} to enhance conventional IM techniques. For example, \cite{8981888} integrates RIS with receive space shift keying (RIS-RSSK) and receive spatial modulation (RIS-RSM). Both schemes encode $\log_2(n_R)$ index bits by selecting a single receive antenna, while the RIS uses CSI for passive beamforming to direct signal power to the chosen antenna, conveying the information bits. In~\cite{Yuan2021,10214326}, receive quadrature spatial modulation (RQSM) is employed in RIS-assisted systems to enhance SE over RIS-RSM by mapping I/Q components to different antennas, effectively doubling index bits. For example,~\cite{10214326} proposes RIS-RQRM, partitioning the RIS into two sub-surfaces for separate I/Q dimensions.

In \cite{Marin2024}, RIS is integrated with receive generalized SSK (RIS-RGSSK) and receive generalized SM (RIS-RGSM), extending the RIS-RSSK and RIS-RSM paradigms by enabling the selection of multiple receive antennas to convey index bits. Other works, including~\cite{9698029,10388479,10416245,10685086}, investigate RIS-assisted spatial or index modulation schemes under the assumption of perfect CSI at the RIS. Specifically,~\cite{9698029,10388479} analyze RIS-RSSK and RIS-RSM with greedy detection, deriving closed-form error probability expressions and highlighting performance saturation at extreme SNRs. In~\cite{10416245}, joint active and passive beamforming is introduced for RIS-RSM to enhance antenna index detection. Meanwhile,~\cite{10685086} proposes antenna selection strategies within RIS-RSM systems to improve SE and reduce detection complexity. To eliminate CSI overhead at the receiver, the RIS-empowered differential chaos shift keying scheme with nested IM was proposed in~\cite{10844041}. While the receiver in~\cite{10844041} remains non-coherent, the RIS still requires channel-phase knowledge to align its reflections and maximize the received power.

\textit{RIS as information encoder:}  
In this category, the RIS acts as an information encoder by implementing IM via its reflection patterns \cite{9217944,9729740,10729864,10623435,10547354,10257609,9838719,10138929,9965423,10693281,10937504}; when CSI is available at the RIS---albeit challenging---it can additionally align signals constructively at the receiver to boost performance. Spatial modulation (SM) and generalized SM (GSM) have been widely adapted to RIS (RIS-SM/GSM) \cite{9217944,10729864}, where information is encoded via the indices of one or more active RIS element groups—analogous to their use in the antenna domain.

In \cite{9217944}, RIS-GSM based scheme is proposed which is termed as reflection pattern modulation (RPM) \footnote{RPM is a general term used in literature to encompass all IM schemes that utilize RIS reflection patterns to encode information.}, with RIS elemnts phase shifts optimized to enhance received power. In \cite{10729864}, the authors consider MISO-RIS system and combine the concepts of RPM  and power distribution index modulation (PIM) to convey information via jointly designed reflection patterns and power allocation across RIS elements, enhancing SE and error performance under full CSI availability at RIS.

Another approach treats the RIS as a multiple-input, multiple-output system (RIS-MIMO) \cite{9535453}, where individual reflecting elements or distinct element groups function as $M$-ary PSK modulators (also known as reflection phase modulators) when illuminated by an unmodulated carrier . In \cite{9729740}, the concept of RIS-MIMO is combined with transmitter-side index modulation based on space-shift keying, whereas the RIS is utilized as a virtual $M$-ary PSK modulator transmitting independent parallel data streams.

In \cite{10623435}, the conventional concept of number modulation and RIS-MIMO is leveraged to introduce a joint number and phase modulation (NPM), where CSI at the RIS is a critical component of the system. Moreover, rectangular differential reflecting spatial modulation (RDRSM)~\cite{10547354} and augmented pattern index modulation (APIM)~\cite{10257609,9838719} jointly map information to RIS reflection patterns and transmitter/receiver indices. In \cite{Liu2023}, an MSIO-based RIS system is considered, where the $K$-means algorithm is utilized to optimize the reflection pattern at the RIS to maximize the received power for a small number of RIS elements. However, this approach quickly becomes infeasible as the number of RIS elements increases.  Hybrid active/passive RIS architectures address the double path loss limitation of fully passive systems. In hybrid reflection modulation (HRM), active elements with power amplifiers encode extra bits \cite{9965423,10693281,10937504}, improving signal strength at the cost of increased hardware complexity and reduced power efficiency. The RIS-CIM scheme proposed in~\cite{10138929} utilizes spreading code indices to convey additional information alongside conventional $M$-ary symbols, enabled by an RIS as a passive beamformer—albeit with increased bandwidth demands due to spreading code usage.

\subsection{Motivation and contribution}
 
One of the most widely used IM principles in RIS is generalized spatial modulation (RIS-GSM) as discussed previously, where the RIS is divided into smaller subsurfaces and information is conveyed through the indices of activated subsurfaces. For example, if the RIS is partitioned into $N_Q$ groups, the maximum number of bits is achieved when exactly $N_Q/2$ groups are activated and the remaining $N_Q/2$ are deactivated. For instance, with $N_Q = 8$, only $\lfloor \log_2 \binom{8}{4} \rfloor=6$ bits  can be transmitted, despite the large number of partitions. Moreover, implementing ultra-fast control to dynamically activate or deactivate subsurfaces according to incoming bits---especially for large RIS panels---can be challenging, often requiring fewer subsurfaces in practice. This reduction not only lowers spectral efficiency but also causes power loss due to the deactivation of half of the reflecting elements.  

Motivated by these limitations and by viewing the RIS as an information encoder, this work proposes a new \textit{RIS-enabled channel signature modulation} (RIS-CSM) scheme for a single-antenna transmitter and a multiple-antenna receiver. RIS-CSM offers a more flexible IM framework capable of achieving higher spectral efficiencies. The scheme encodes information into the indices of $K$ orthogonal binary RIS phase-shift patterns, selected from the rows (or columns) of a large-dimensional Hadamard matrix. Unlike conventional methods that rely on passive or active beamforming at the RIS, RIS-CSM performs no RIS-side beamforming and does not require CSI at the RIS, instead focusing entirely on enhancing spectral efficiency. The RIS elements are partitioned into $N_Q$ disjoint groups, each undergoing CSM processing to generate distinct effective channel signatures, which are then used as indices for the transmitted bits.
 
  A summary of main contributions of this work are as follows:

\begin{itemize}[labelindent=0pt, leftmargin=*]
\item We propose an RIS-assisted SIMO modulation scheme (RIS-CSM) in which each RIS group selects one of $K$ orthogonal binary phase-shift patterns ($0$ or $\pi$), creating distinct channel signatures for data embedding. The group- and pattern-level configuration offers flexible bit-to-pattern mapping, enabling scalable spectral efficiency.

\item A closed-form upper bound on the error probability under ML detection is derived for independent and identically distributed (IID) Rayleigh fading channels, showing close agreement with simulation results.

\item Asymptotic and capacity analyses show the system attains diversity order $n_R$ with coding gain scaling linearly with the number of RIS elements.

\item Extensive simulations are conducted to validate the theoretical analysis and to evaluate the performance of RIS-CSM under a wide range of configurations. 
Comparisons are made with representative baseline schemes—RIS-MIMO, RIS-GSM, and RIS-CIM—that align with our design philosophy in avoiding RIS-side beamforming, employing optimal ML detection, and utilizing the RIS as an information encoder. 
All schemes are evaluated under identical spectral efficiency and system constraints. 
The results indicate that RIS-CSM offers a promising and competitive alternative, achieving favorable BER performance while preserving a simple modulation mechanism.

\item The impact of spatial correlation on RIS-CSM is investigated by extensive simulations, revealing a fundamental trade-off between SNR enhancement and channel diversity loss. For low spectral efficiency (small number of phase-shift patterns), moderate correlation can improve detection and capacity compared to the IID case, while for higher spectral efficiency (large number of phase-shift patterns), the IID channel offers superior performance. These trends are consistent across both BER and capacity results, highlighting the intricate interplay between correlation and system configuration.The performance improvement observed with correlation is consistent with the findings reported in prior work~\cite{Guo2023}.

\end{itemize}

\subsection{Key features of the proposed scheme}
\label{subsec:features}
\begin{itemize}[labelindent=0pt, leftmargin=*]
    \item \textit{No RIS-side beamforming:} The primary goal is to enhance spectral efficiency; hence, no RIS-side beamforming is performed, and CSI is not required at the RIS. CSI is needed only at the receiver for detection, where only $K N_Q$ effective channel signatures are estimated—significantly fewer than the total number of individual channel links.  

    \item \textit{Efficient Pattern Generation:} Reflection patterns derived from a Hadamard matrix can be generated efficiently either all at once, stored, or generated one by one on the fly~\cite{Monroy2024}. This approach operates with simple binary phase control hardware at RIS.

    \item \textit{Flexible Bit-to-Pattern Mapping:} There is a straightforward bit-to-reflection pattern mapping per group, where index bits are converted to a decimal value used to select the corresponding phase-shift pattern from a predefined set. All groups share the same \(K\) orthogonal patterns, with flexible tradeoffs in \(K\) and \(N_Q\) enabling a wide range of spectral efficiencies.

    \item \textit{Manageable ML Detection Complexity:} The receiver uses ML detection with complexity \(\mathcal{O}(K^{N_Q} n_R)\). For typical values of \(K\) and \(N_Q\), this complexity remains practical thanks to the tradeoff between these parameters, which enables a flexible range of spectral efficiencies. For instance, in a system with a square $8\times 8$ RIS, configurations $(N_Q, K, n_R) = (1, 8, 1)$ and $(1, 16, 1)$ yield spectral efficiencies of 3 and 4 bits, requiring 8 and 16~ML evaluations, respectively. Increasing to $(2, 8, 1)$ and $(2, 16, 1)$ doubles the spectral efficiencies to 6 and 8 bits, with 64 and 256~ML evaluations, respectively---still computationally feasible.
        
   \item \textit{High Spectral Efficiency:} The proposed scheme achieves 
$R = \log_2(M) + N_Q \log_2(K)$ bits per channel use (bpcu) when $M$-ary modulation is employed at the transmitter, where $M$ is the modulation order. For example, with an $8 \times 8$ RIS, $M=4$, $N_Q=2$, and $K=16$, the resulting SE is $R = 2 + 8 = 10$~bpcu.

\end{itemize}

The remainder of this paper is organized as follows: Section~\ref{sec2} presents the system model and RIS-CSM scheme. Section~\ref{sec3} analyzes performance through a derived closed-form error probability bound. Section~\ref{sec4} examines asymptotic error behavior and the impact of RIS partitioning, including capacity analysis. Section~\ref{sec5} provides simulation results, and Section~\ref{sec6} concludes with future research directions.

\textbf{Notation:} In this work, bold lowercase letters such as $\mathbf{x}$ represent vectors, bold uppercase letters such as $\mathbf{X}$ represent matrices, non-bold letters such as \(x\) represent scalars, \(\mathbf{A} = \operatorname{diag}(\mathbf{a})\) denotes a diagonal matrix with the entries of \(\mathbf{a}\) on its diagonal, statistical expectation is denoted by \(\mathbb{E}[\cdot]\), the transpose of a matrix or vector is denoted by \((\cdot)^T\), the Hermitian transpose is denoted by \((\cdot)^H\), and the binomial coefficient is denoted by \(\binom{n}{k}\), representing the number of ways to choose \(k\) items from \(n\) items without regard to order.

\section{System model}
\label{sec2}
 
We consider a dual-hop communication system \cite{canbilen2020reconfigurable, 9729740} comprising a single-antenna transmitter (Tx), an RIS with \(N\) reflecting elements, and an \(n_R\)-antenna receiver (Rx), with the RIS located in the far-field of both terminals. The RIS in our case can operate as a backscatterer or be remotely transmitter-controlled via a secure link. Due to a blocked direct link, communication occurs solely through the RIS-assisted path.

Assuming a rich-scattering environment with multipath propagation and no line-of-sight (LOS) paths from Tx to RIS and from RIS to Rx, the channels are modeled as narrowband flat-fading. Under these conditions, let \(\mathbf{h} \in \mathbb{C}^N\) denote the Tx-RIS small-scale fading channel and \(\mathbf{G} \in \mathbb{C}^{n_R \times N}\) represent the RIS-Rx small-scale fading channel. Both channels comprise IID complex Gaussian random variables with zero mean and unit variance.

Furthermore, it is assumed that the RIS array is compact relative to the link distances, such that all RIS elements experience approximately uniform path loss from the transmitter, and each RIS element has similar path loss to all receiver antennas \cite{canbilen2020reconfigurable, 9729740}. These path loss effects are implicitly absorbed into the system SNR definition, allowing us to focus on the fading characteristics while maintaining analytical tractability.

\subsection{Proposed RIS-CSM system}
\begin{figure*}[htp!]
	\centering
	\includegraphics[width=0.7\textwidth]{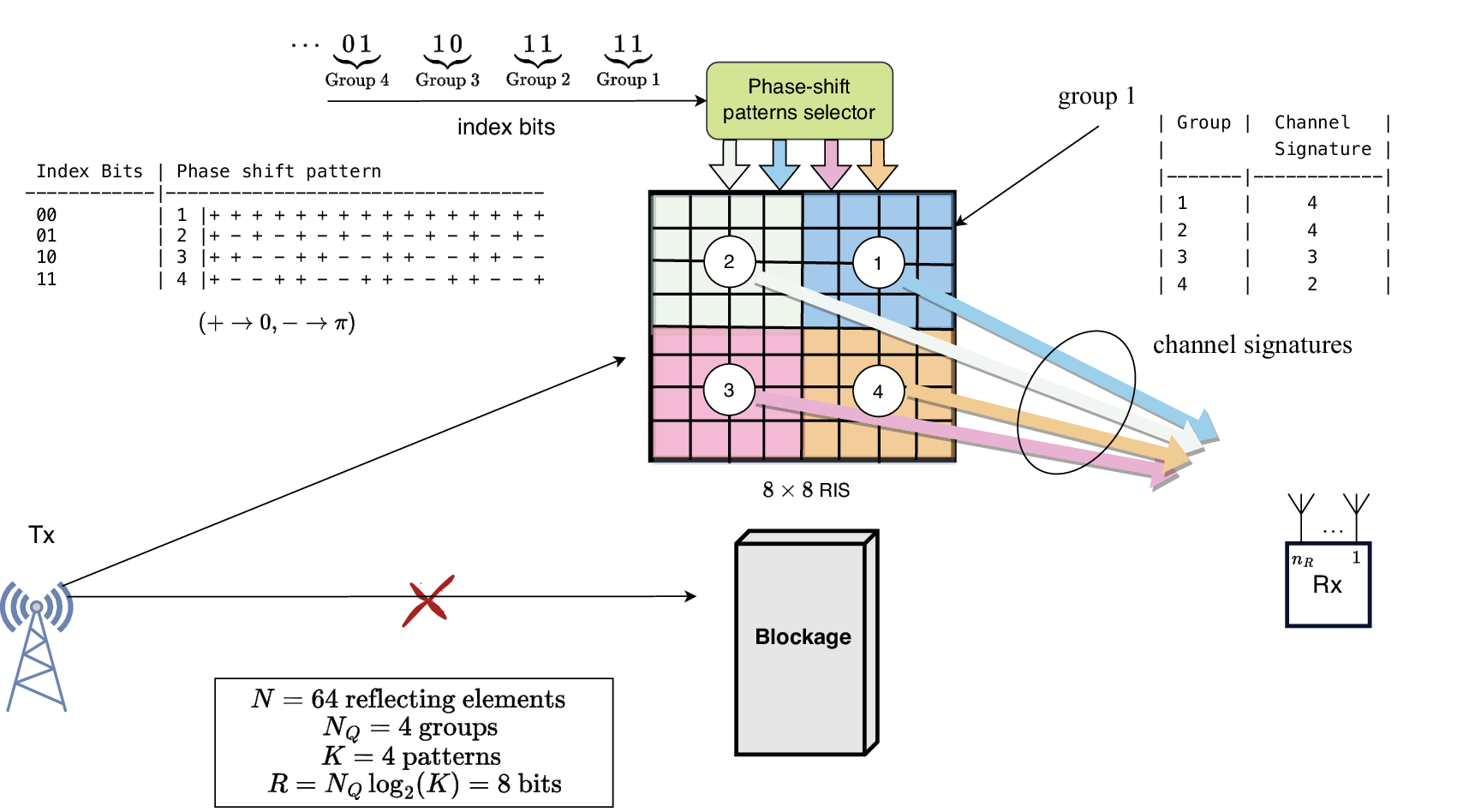}
\caption{Conceptual schematic of the RIS-CSM system in a SIMO RIS-assisted setup with a square \(8 \times 8\) RIS divided into \(N_Q=4\) groups, each containing \(n=16\) elements. The RIS operates either as a backscatterer or under transmitter control in Rayleigh fading. For each group, \(\log_2(K)\) index bits are mapped to one of \(K=4\) phase-shift patterns derived from a \(16 \times 16\) Hadamard matrix. Each reflection pattern produces a distinct effective channel signature at the receiver, illustrated with different colors. In total, \(4\log_2(4)=8\) bits are embedded across the four group signatures. The receiver recovers the transmitted bits by identifying the tuple of four channel signatures.}

	\label{fig:fig1}
\end{figure*}

In RIS-CSM, the RIS elements are partitioned into \(N_Q\) disjoint groups, with each group containing $n = N/N_Q$ reflecting elements. For the $q$-th group ($q \in \{1, \dots, N_Q\}$), the Tx-RIS and RIS-Rx channels are denoted by $\mathbf{h}_q \in \mathbb{C}^n$ and $\mathbf{G}_q \in \mathbb{C}^{n_R \times n}$, respectively, where  $\mathbf{h}_q$ is the sub-vector of $\mathbf{h}$ corresponding to the indices of elements in group $q$, and $\mathbf{G}_q$ is the sub-matrix of $\mathbf{G}$ comprising the columns corresponding to the indices of elements in group $q$, e.g., $\mathbf{h} = [\mathbf{h}_1^T \ \mathbf{h}_2^T \ \dots \ \mathbf{h}_{N_Q}^T]^T$ and $\mathbf{G} = [\mathbf{G}_1 \ \mathbf{G}_2 \ \dots \ \mathbf{G}_{N_Q}]$. Fig.~\ref{fig:fig1} illustrates a conceptual schematic diagram of the system, with the RIS partitioned into four distinct groups.

The information is encoded by selecting one of $K\leq n$ RIS phase-shift configurations (phase-shift vectors) for each group from the set $\mathcal{S}=\{\mathbf{s}_1, \mathbf{s}_2, \dots, \mathbf{s}_K\}$. We consider $\mathcal{S}$ to be the first $K$ rows/columns of an $n \times n$ Hadamard matrix $\mathbf{H}_n$, where $K = 2^i$ for some integer $i$. Consequently, each reflecting element in the RIS is configured with a phase shift of either $0$ or $\pi$, while maintaining a constant signal amplitude. When the transmitter employs no conventional M-ary modulation, the SE of the system is:
\begin{equation}
	R = N_Q \log_2 K \quad \text{bpcu}.
\end{equation}

The transmitter groups the incoming bit stream into blocks of $R$ bits. These blocks are divided into $N_Q$ index sub-blocks of $\log_2(K)$ bits each, used to select the phase-shift configuration for each RIS group, and, when M-ary modulation is used, a modulation sub-block of $\log_2(M)$ bits is used to select the transmitted modulation symbol. In this case $R=\log_2(M)+N_Q\log_2(K)$ bpcu.

Table~\ref{tab:example1} provides an example for an RIS with $N = 16$ elements and a single group ($N_Q=1$), where the first $K = 4$ rows of the Hadamard matrix $\mathbf{H}_{16}$ are mapped to bit sequences.

\begin{table*}[htp!]
	\centering
	\caption{Bit assignment for RIS with $N = 16, N_Q=1, K = 4$.}
	\label{tab:example1}
	\begin{tabular}{|c|c|l|}
		\hline
		\textbf{IM bits} & \multicolumn{1}{c|}{\textbf{Reflection pattern}} \\ \hline
		00 & \(+1, +1, +1, +1, +1, +1, +1, +1, +1, +1, +1, +1, +1, +1, +1, +1\) \\ \hline
		01 & \(+1, -1, +1, -1, +1, -1, +1, -1, +1, -1, +1, -1, +1, -1, +1, -1\) \\ \hline
		10 & \(+1, +1, -1, -1, +1, +1, -1, -1, +1, +1, -1, -1, +1, +1, -1, -1\) \\ \hline
		11 & \(+1, -1, -1, +1, +1, -1, -1, +1, +1, -1, -1, +1, +1, -1, -1, +1\) \\ \hline
	\end{tabular}
\end{table*}

Let \(\mathbf{k} = [k_1, \ldots, k_{N_Q}] \in \mathcal{K}\) be the index vector that specifies the phase-shift configuration selection for all groups, where $\mathcal{K} = \{1, \ldots, K\}^{N_Q}$ is the set of all possible indices with cardinality $|\mathcal{K}|=K^{N_Q}$. Here, each \(k_q \in \{1, \ldots, K\}\) denotes the chosen phase-shift pattern for group \(q\). For a given \(\mathbf{k}\), the received signal \(\mathbf{y}\) can be expressed as:
\begin{equation}
	\label{eq:rx_signal}
	\mathbf{y} = \sqrt{E_s} \sum_{q=1}^{N_Q} \underbrace{\mathbf{G}_q \operatorname{diag}(\mathbf{h}_q) \mathbf{s}_{k_q}}_{\mathbf{d}_{k_q}} x + \mathbf{n}
\end{equation}
where $\mathbf{d}_{k_q}$ is the effective channel of $q$-th group, $x$ represents the transmitted symbol, assumed to have normalized average power and $\mathbf{n} \in \mathbb{C}^{n_R}$ is additive complex Gaussian noise with zero mean and covariance matrix $\operatorname{E}\{\mathbf{n} \mathbf{n}^H\} = \sigma^2 \mathbf{I}_{n_R}$. Here, \(\mathbf{h}_q \in \mathbb{C}^n\) denotes the channel vector from the transmitter to the \(q\)-th RIS group. The operator \(\operatorname{diag}(\mathbf{h}_q)\) creates an \(n \times n\) diagonal matrix whose diagonal elements are the entries of \(\mathbf{h}_q\).

In the sequel, unless stated otherwise, we focus on the system without $M$-ary modulation, where the transmitter sends an unmodulated carrier (i.e., $x = 1$). From~\eqref{eq:rx_signal}, the overall effective channel seen by the receiver is defined by:
\begin{equation}
	\mathbf{d}_{\mathbf{k}}=\sum_{q=1}^{N_Q} \mathbf{d}_{k_q} =\sum_{q=1}^{N_Q} \mathbf{G}_q \operatorname{diag}(\mathbf{h}_q) \mathbf{s}_{k_q}
	\label{eq:effective_channel_GCSM}
\end{equation}

The effective channel $	\mathbf{d}_{\mathbf{k}}$ is uniquely determined by the $N_Q$ RIS phase-shift vectors $\mathbf{s}_{k_q}$ ($q = 1, \ldots, N_Q$), resulting in $K^{N_Q}$ distinct channel signatures. From \eqref{eq:rx_signal}, the signal-to-noise ratio (SNR) is defined as $\mathrm{SNR} = E_s/\sigma^2$.

Assuming the effective channels $\mathbf{d}_{\mathbf{k}}$ ($\mathbf{k} \in \mathcal{K}$) are known at the receiver, the maximum likelihood (ML) detection rule for recovering the transmitted index vector \(\mathbf{k}\) is formulated as:
\begin{equation}
	\label{eq:ML}
	\hat{\mathbf{k}} = \arg\min_{\mathbf{k} \in \mathcal{K}} \left\| \mathbf{y} - \sqrt{E_s} \mathbf{d}_{\mathbf{k}} \right\|^2.
\end{equation}

The ML requires computing $K^{N_Q}$ Euclidean distances of dimension $n_R$, yielding a computational complexity of $\mathcal{O}(K^{N_Q} n_R)$.

\subsection{Properties and generation of phase-shift patterns}A Hadamard matrix of order $n$, denoted $H_n$, is an $n \times n$ matrix with entries in $\{+1, -1\}$ satisfying $H_n^\top H_n = H_n H_n^\top = n I_n$. A necessary condition for the existence of a Hadamard matrix is that its order is $n = 1$, $2$, or a multiple of 4, with mutually orthogonal rows and columns~\cite{Hedayat1999}.

As described earlier, the $q$-th group of reflecting elements is configured using the first $K$ rows or columns of an $n \times n$ Hadamard matrix $H_n$, where $N$, $K$, and $n = N / N_Q$ are powers of two. This enables the use of Sylvester's recursive construction for Hadamard matrices via the Kronecker product~\cite{Hedayat1999}:
\begin{equation}
\label{eq:Sylvesterformula}
    H_{n} = H_2 \otimes H_{n/2}, \quad \text{with } H_2 = \begin{pmatrix} 1 & 1 \\ 1 & -1 \end{pmatrix},
\end{equation}
which enables orthogonal phase-shift patterns at RIS using simple 1-bit ($0$ or $\pi$) modulators, which are far less complex and costly than analog or multi-bit phase shifters.
 
It is worth noting that by exploiting the grouped Kronecker form $H_{n} = H_K \otimes H_{n/K}$---which generalizes the formula in \eqref{eq:Sylvesterformula}---the first $K$ rows of $H_n$ are formed by taking each row of $H_K$ and repeating it $n/K$ times as a contiguous block. This is a direct consequence of the Kronecker product structure and the recursive nature of Hadamard matrices. For example, consider generating the first $K = 8$ rows of a $256 \times 256$ Hadamard matrix. First, an $8 \times 8$ Hadamard matrix is generated via Sylvester's recursion, and each of its rows is repeated $256/8 = 32$ times to form the first $K$ rows of $H_{256}$. Alternatively, to efficiently generate specific rows on-the-fly at the receiver, the Hadamard row-wise generation algorithm proposed in~\cite{Monroy2024} can be used, which requires almost no storage overhead.

Let the columns of $\mathbf{H}_n$ be denoted as $\mathbf{s}_1, \mathbf{s}_2, \ldots, \mathbf{s}_n$. Their orthogonality is given by:
\begin{equation}
	\mathbf{s}_m^T \mathbf{s}_\ell =
	\begin{cases}
		n, & \text{if } m = \ell, \\
		0, & \text{if } m \neq \ell.
	\end{cases}
	\label{eq:orthogonality}
\end{equation}

Additionally, for any distinct columns $\mathbf{s}_m$ and $\mathbf{s}_\ell$ ($m \neq \ell$), the element-wise difference $\Delta_{m\ell} = \mathbf{s}_m - \mathbf{s}_\ell$ yields a vector with half of its components equal to $\pm 2$ and the other half equal to zero \cite{ProakisManolakis2007}:
\begin{equation}
	\Delta_{m\ell,j} =
	\begin{cases}
		\pm 2, & \text{for } n/2 \text{ components}, \\
		0, & \text{for } n/2 \text{ components},
	\end{cases}
	\label{eq:H_property}
\end{equation}
where $\Delta_{m\ell,j}$ is the $j$-th entry of $\Delta_{m\ell}$.

\subsection{Channel estimation}
Since the ML detector~\eqref{eq:ML} directly depends on the effective channel \(\mathbf{d}_{\mathbf{k}}\), which incorporates the phase-shift vectors, we focus on estimating \(\mathbf{d}_{\mathbf{k}}\) directly. Specifically, we estimate the effective channel \(\mathbf{d}_{k_q}\) for the $q$-th group (\(k_q \in \{1, 2, \ldots, K\}\)) in a round-robin fashion, with all other groups deactivated. Each group requires at least \(K\) measurements, resulting in a total of \(N_Q K\) measurements across all \(N_Q\) groups. This is typically much smaller than \(N\). For example, in a 4-bit system with \(N = 128\), \(N_Q = 2\), and \(K = 4\), the total measurements are \(N_Q K = 8 \ll 128\).

During the training phase, the transmitter sends a constant symbol \(x = 1\) with power \(E_t\) over \(\tau\) symbol intervals, while the RIS applies the phase-shift vector \(\mathbf{s}_{k_q}\). This yields \(\tau\) noisy observations:
\begin{equation}
	\mathbf{z}_i = \sqrt{E_t} \mathbf{d}_{k_q} + \mathbf{w}_i, \quad i = 1, 2, \ldots, \tau,
	\label{eq:observation_model}
\end{equation}
where \(\mathbf{z}_i \in \mathbb{C}^{n_r}\) is the received signal, \(\mathbf{d}_{k_q} = \mathbf{G}_q \operatorname{diag}(\mathbf{h}_q) \mathbf{s}_{k_q}\), and \(\mathbf{w}_i \sim \mathcal{CN}(\mathbf{0}, \mathbf{I}_{n_r})\) is normalized complex Gaussian noise.

Since the observations are IID due to uncorrelated Gaussian noise samples, the sample mean is a sufficient statistic:
\begin{equation}
	\bar{\mathbf{z}} = \frac{1}{\tau} \sum_{i=1}^\tau \mathbf{z}_i = \sqrt{E_t} \mathbf{d}_{k_q} + \bar{\mathbf{w}},
	\label{eq:sample_mean}
\end{equation}
where \(\bar{\mathbf{w}} = \frac{1}{\tau} \sum_{i=1}^\tau \mathbf{w}_i\) has covariance \(\frac{1}{\tau} \mathbf{I}_{n_r}\). Using \eqref{eq:sample_mean}, the minimum mean square error (MMSE) estimate is derived as:
\begin{equation}
	\hat{\mathbf{d}}_{k_q} = \mathbf{C}_{\mathbf{d}_{k_q} \bar{\mathbf{z}}} \mathbf{C}_{\bar{\mathbf{z}}}^{-1} \bar{\mathbf{z}},
	\label{eq:mmse_estimator}
\end{equation}
with covariance terms:
\begin{align}
	\mathbf{C}_{\mathbf{d}_{k_q} \bar{\mathbf{z}}} &= \frac{\sqrt{E_t} N}{N_Q} \mathbf{I}_{n_r}, \label{eq:cross_cov} \\
	\mathbf{C}_{\bar{\mathbf{z}}} &= \left( \frac{E_t N}{N_Q} + \frac{1}{\tau} \right) \mathbf{I}_{n_r}. 
	\label{eq:obs_cov}
\end{align}

Substituting \eqref{eq:cross_cov} and \eqref{eq:obs_cov} into \eqref{eq:mmse_estimator} yields:
\begin{equation}
	\hat{\mathbf{d}}_{k_q} = \frac{\sqrt{E_t} N}{E_t N \tau + N_Q} \sum_{i=1}^\tau \mathbf{z}_i, 
	\label{eq:mmse_final}
\end{equation}
and from the orthogonality principle of MMSE, the MSE for each estimated component of \(\hat{\mathbf{d}}_{k_q}\) is given by
\begin{equation}
	\mathrm{MSE}= \frac{N}{E_t N \tau + N_Q}.
	\label{eq:MSE}
\end{equation}

\begin{figure}[t]
	\centering
	\includegraphics[width=0.45\textwidth]{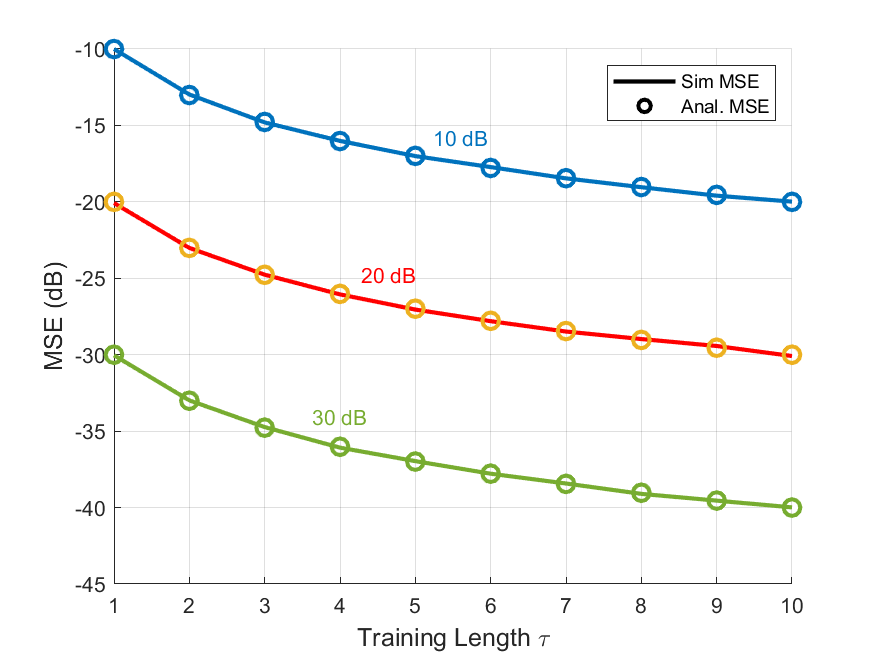}
	\caption{MSE of the estimated effective channel for a system with \(N = 128\) and \(N_Q = 1\). }
	\label{fig:fig2}
\end{figure}

Fig.~\ref{fig:fig2} shows the MSE performance for various training SNR values and training lengths, \(\tau\) with $N=128$ and $N_Q=1$. As expected, MSE decreases with higher SNR and longer \(\tau\). The figure reveals a trade-off between SNR and $\tau$ suggesting that higher training SNRs can potentially reduce the required training overhead which is important in RIS-assisted systems.

\section{Performance analysis}
\label{sec3}

The ML detector in \eqref{eq:ML} detects the supersymbol \(\hat{\mathbf{k}} = [\hat{k}_1, \dots, \hat{k}_{N_Q}]\), where an error occurs if \(\mathbf{k} \neq \hat{\mathbf{k}}\). However, this does not reflect the per-group error probability, particularly in low SNR regimes where multiple group errors are highly likely. Thus, the subsequent analysis focuses on the average symbol error rate (SER) for an individual group, specifically the error probability of the per-group index symbol.

Given the index vector $\mathbf{k} \in \mathcal{K}$ was transmitted and $\hat{\mathbf{k}} \neq \mathbf{k} $ is detected, the average SER can be approximated using the union bound \cite{ProakisManolakis2007} \cite{SimonAlouini2005}:
\begin{equation}
	P_e \leq \frac{1}{N_Q |\mathcal{K}|} \sum_{\mathbf{k} \in \mathcal{K}} \sum_{\substack{\hat{\mathbf{k}} \in \mathcal{K} \\ \hat{\mathbf{k}} \neq \mathbf{k}}} \bar{P}_{\mathbf{k} \hat{\mathbf{k}}} \ell({\mathbf{k}},\hat{\mathbf{k}}),
	\label{eq:unionbound}
\end{equation}
where $\bar{P}_{\mathbf{k}\hat{\mathbf{k}}} \triangleq \mathbb{E}[P_{\mathbf{k}\hat{\mathbf{k}}}]$ is the average pairwise error probability (PEP) over channel realizations, with $P_{\mathbf{k}\hat{\mathbf{k}}} \triangleq \Pr[\mathbf{k} \to \hat{\mathbf{k}}]$ denoting the conditional PEP, conditioned implicitly on all channel randomness. The function $\ell(\mathbf{k}, \hat{\mathbf{k}})$ represents the Hamming distance between the indices $\mathbf{k}$ and $\hat{\mathbf{k}}$, defined as:
\begin{equation}
	\ell(\mathbf{k}, \hat{\mathbf{k}}) = \sum_{q=1}^{N_Q} \mathbb{I}(k_q \neq \hat{k}_q),
\label{eq:EuclideanDis}
\end{equation}
where \(\mathbf{k} = [k_1, \ldots, k_{N_Q}] \in \mathcal{K}\), \(\hat{\mathbf{k}} = [\hat{k}_1, \ldots, \hat{k}_{N_Q}] \in \mathcal{K}\), and \(\mathbb{I}(\cdot)\) is the indicator function, equal to 1 if \(k_q \neq \hat{k}_q\) and 0 otherwise.

The conditional PEP can be expressed as:
\begin{equation}
	P_{\mathbf{k} \hat{\mathbf{k}}} = \operatorname{Pr} \left[ \|\mathbf{y} - \sqrt{E_s} \mathbf{d}_{\mathbf{k}}\|^2 > \|\mathbf{y} - \sqrt{E_s} \mathbf{d}_{\hat{\mathbf{k}}}\|^2 \right],
	\label{eq:PEP1}
\end{equation}
where, by substituting \(\mathbf{y}\) from the channel model \eqref{eq:rx_signal} and performing algebraic manipulations, \eqref{eq:PEP1} reduces to:
\begin{equation}
	P_{\mathbf{k} \hat{\mathbf{k}}} = \operatorname{Pr} \left[ 2 \sqrt{E_s} \operatorname{Re} \{ \mathbf{n}^H (\mathbf{d}_{\mathbf{k}} - \mathbf{d}_{\hat{\mathbf{k}}}) \} < -E_s \|\mathbf{d}_{\mathbf{k}} - \mathbf{d}_{\hat{\mathbf{k}}}\|^2 \right].
	\label{eq:PEP2}
\end{equation}

Since $\mathbf{n}$ is complex Gaussian with mean 0 and covariance matrix $\sigma^2 \mathbf{I}_{n_R}$, the term $z = 2 \sqrt{E_s} \operatorname{Re}\{\mathbf{n}^H (\mathbf{d}_{\mathbf{k}} - \mathbf{d}_{\hat{\mathbf{k}}})\}$ is a real Gaussian random variable with mean zero and variance $\operatorname{Var}(z) = 2 \sigma^2 E_s \| \mathbf{d}_{\mathbf{k}} - \mathbf{d}_{\hat{\mathbf{k}}} \|^2$. As such, \eqref{eq:PEP2} can be simplified as
\begin{equation}
	P_{\mathbf{k} \hat{\mathbf{k}}} = Q\left( \sqrt{\frac{E_s \|\mathbf{d}_{\mathbf{k}} - \mathbf{d}_{\hat{\mathbf{k}}}\|^2}{2 \sigma^2}} \right),
	\label{eq:Conditional_Error_Prob2}
\end{equation}

The effective channel difference for an error event (\(\hat{\mathbf{k}} \neq \mathbf{k}\)) is defined as:
\begin{align}
	\mathbf{d}_{\mathbf{k}} - \mathbf{d}_{\hat{\mathbf{k}}} &= \sum_{q\in \mathcal{D}} \mathbf{G}_q \operatorname{diag}(\mathbf{h}_q) (\mathbf{s}_{k_q} - \mathbf{s}_{\hat{k}_q}) \notag \\
	&= \sum_{q \in \mathcal{D}} \mathbf{G}_q \operatorname{diag}(\mathbf{h}_q) \Delta_{k_q \hat{k}_q},
	\label{eq:differing_indices}
\end{align}
where \(\Delta_{k_q \hat{k}_q} = \mathbf{s}_{k_q} - \mathbf{s}_{\hat{k}_q}\), and \(\mathcal{D} = \{ q : k_q \neq \hat{k}_q \}\) denotes the set of groups with differing indices, with \(|\mathcal{D}|\) representing the Hamming distance. When \(k_q = \hat{k}_q\), the term \(\mathbf{G}_q \operatorname{diaq}(\mathbf{h}_q) (\mathbf{s}_{k_q} - \mathbf{s}_{\hat{k}_q}) = \mathbf{0}\), contributing nothing to the difference.

For each \(q \in \mathcal{D}\), define:
\begin{equation}
	\mathbf{z}_q = \mathbf{G}_q \operatorname{diag}(\mathbf{h}_q) \Delta_{k_q \hat{k}_q}.
\end{equation}
The \(i\)-th component of \(\mathbf{z}_q\) is:
\begin{equation}
	z_{q,i} = \sum_{j=1}^n [\mathbf{G}_q]_{i,j} h_{q,j} \Delta_{k_q \hat{k}_q,j}
	\label{eq:z_q}
\end{equation}
where \([\mathbf{G}_q]_{i,j}\) and \(h_{q,j}\) are the \((i,j)\)-th entry of \(\mathbf{G}_q\) and the \(j\)-th entry of \(\mathbf{h}_q\), respectively, both distributed as $\mathcal{CN}(0,1)$. 

From \eqref{eq:H_property}, $ \Delta_{k_q \hat{k}_q}$ has exactly $n/2$ nonzero entries of value $\pm 2$ (for $j \in \mathcal{V}$) and zeros elsewhere, where $\mathcal{V}$ denotes the set of indices with nonzero differences ($|\mathcal{V}| = n/2$). This allows rewriting \eqref{eq:z_q} as
\begin{equation}
	z_{q,i}  = 2 \sum_{j \in \mathcal{V}} [\mathbf{G}_q]_{i,j} h_{q,j} \operatorname{sign}(\Delta_{k_q \hat{k}_q,j})
\end{equation}
where $\operatorname{sign}(\cdot)$ denotes the sign operator. The product \([\mathbf{G}_q]_{i,j} h_{q,j}\) admits a zero mean and unit variance random variable, and for large $n$, the sum approximates a Gaussian distribution by the virtue of CLT. Since the distribution is invariant to the sign of $\Delta_{k_q \hat{k}_q,j}$, this formulation shows that each component of $\mathbf{z}_q$ is a sum of $n/2$ random variables, each with zero mean and variance
\begin{equation}
	\text{Var}(z_{q,i}) = \sum_{j \in \mathcal{V}} | \Delta_{k_q \hat{k}_q,j} |^2 = \|\Delta_{k_q \hat{k}_q}\|^2 = 2n,
\end{equation}
yielding $\mathbf{z}_q \sim \mathcal{CN}(0, 2n \mathbf{I}_{n_R})$ as an approximation by CLT.

Since the \(\mathbf{z}_q\) terms are independent across groups, the total channel difference is:
\begin{equation}
	\mathbf{d}_{\mathbf{k}} - \mathbf{d}_{\hat{\mathbf{k}}} = \sum_{q \in \mathcal{D}} \mathbf{z}_q   \sim \mathcal{CN}(0, 2n |\mathcal{D}| \mathbf{I}_{n_R}),
\end{equation}
and hence the squared norm is:
\begin{equation}
	\|\mathbf{d}_{\mathbf{k}} - \mathbf{d}_{\hat{\mathbf{k}}}\|^2 = \sum_{i=1}^{n_R} |a_i|^2,
	\label{eq:chii}
\end{equation}
where \(a_i \sim \mathcal{CN}(0, 2n |\mathcal{D}|)\). Thus, the conditional PEP may be written as:
\begin{equation}
	P_{\mathbf{k} \hat{\mathbf{k}}} = Q\left( \sqrt{\frac{E_s \sum_{i=1}^{n_R} |a_i|^2}{2 \sigma^2}} \right).
	\label{eq:Conditional_Error_Prob3}
\end{equation}

\begin{proposition}
	\label{thm:SymbolErrorBound}
	The per-group average SER in the RIS-CSM system is approximately upper bounded as follows:
	\begin{equation}
		\label{eq:thmbound}
		P_e \lesssim \frac{ K^{N_Q-1}}{K^{N_Q} - 1} \sum_{\zeta=1}^{N_Q} \binom{N_Q}{\zeta} (K-1)^{\zeta+1} F(\zeta),
	\end{equation}
	where $F(\zeta)$ is defined by
	\begin{align}
		F(\zeta) &= \left( \frac{1 - \mu(\zeta)}{2} \right)^{n_R} \sum_{k=0}^{n_R-1} \binom{n_R - 1 + k}{k} \left( \frac{1 + \mu(\zeta)}{2} \right)^k,
	\end{align}
	with $	\mu(\zeta) = \sqrt{\frac{N E_s \zeta /\sigma^2}{2 N_Q + N E_s \zeta/\sigma^2}}.$
\end{proposition}

\begin{proof}
	Let 
	\begin{equation}
		\zeta \triangleq |\mathcal{D}|, \quad \zeta \in \{1,2, \cdots, N_Q\}
		\label{eq:zeta_def}
	\end{equation}
	be a discrete random variable. Accordingly, \eqref{eq:chii} can be be rewritten as 
	
	\begin{equation} 
		\sum_{i=1}^{n_R} | {a}_i|^2 = 2n \zeta \sum_{i=1}^{n_R} |\tilde{a}_i|^2 = 2n \zeta \alpha
	\end{equation}
	where \(\tilde{a}_i \sim \mathcal{CN}(0,1)\) and  \(\alpha \sim \Gamma(n_R, 1)\) (Gamma distribution with shape parameter $n_R$ and rate parameter 1), with probability density function (pdf):
	\begin{equation}
		f(\alpha) = \frac{\alpha^{n_R-1}e^{-\alpha}}{\Gamma(n_R)}, \quad \alpha > 0,
	\end{equation}
	where $\Gamma(n_R) = (n_R-1)!$.
	
	Using \eqref{eq:Conditional_Error_Prob3}, the average PEP decomposes as:
	\begin{equation}
		\bar{P}_{\mathbf{k}\hat{\mathbf{k}}} = \mathbb{E}_{\zeta} \left[ \mathbb{E}_{\alpha} \left[ Q\left( \sqrt{\frac{E_s \zeta n \alpha}{\sigma^2}} \right) \Big| \zeta \right] \right],
		\label{eq:xxxx}
	\end{equation}
	where the inner and outer expectations are with respect to the variables $\alpha$ and $\zeta$, respectively.
	
	Let 
	\begin{equation}
		\bar{\gamma} (\zeta) = \frac{E_s \zeta n}{2\sigma^2} = \frac{E_s \zeta N }{2 \sigma^2 N_Q }.
	\end{equation}
	Then, for a fixed $\zeta$, the inner expectation in \eqref{eq:xxxx} can be expressed as:
	\begin{align}
		\nonumber
		F(\zeta) & \triangleq \mathbb{E}_{\alpha} \left[Q\left(\sqrt{2 \bar{\gamma}(\zeta) \alpha}\right)\Big| \zeta \right] \\
		&= \int_{0}^{\infty} Q\left(\sqrt{2 \bar{\gamma}(\zeta) \alpha}\right) f(\alpha) \, d\alpha \nonumber \\
		&= \frac{1}{\Gamma(n_R)} \int_{0}^{\infty} Q\left(\sqrt{2\bar{\gamma}(\zeta) \alpha}\right) \alpha^{n_R-1} e^{-\alpha} \, d\alpha.
		\label{eq:inner_exp1}
	\end{align}
	
	From \cite{SimonAlouini2005}, the integral in \eqref{eq:inner_exp1} can be simplified by using an alternative form of the $Q$-function and some mathematical manipulations:
	\begin{equation}
		F(\zeta) = \frac{1}{\pi} \int_0^{\pi/2} \left(1 + \frac{\bar{\gamma}(\zeta)}{\sin^2\theta}\right)^{-n_R} d\theta,
		\label{eq:Average_PEP3}
	\end{equation}
	and its closed-form solution \cite{ProakisManolakis2007,SimonAlouini2005} is given by:
	\begin{align}
		\nonumber
		&F(\zeta) =\\
		& \left( \frac{1 - \mu(\zeta)}{2} \right)^{n_R} \sum_{k=0}^{n_R-1} \binom{n_R - 1 + k}{k} \left( \frac{1 + \mu(\zeta)}{2} \right)^k,
		\label{eq:inner_exp2}
	\end{align}
	where, in our case, $\mu(\zeta)$ is defined as:
	\begin{equation}
		\mu(\zeta)= \sqrt{\frac{\bar{\gamma}(\zeta)}{1 + \bar{\gamma}(\zeta)}} = \sqrt{\frac{N E_s \zeta /\sigma^2}{2 N_Q+ N E_s \zeta /\sigma^2}}.
		\label{eq:expression_mu}
	\end{equation}
	
	%
	
	
	Now, assume $\hat{\mathbf{k}}$ is a random vector uniformly distributed over $\{1, \ldots, K\}^{N_Q}$. Recall $\mathcal{D} = \{ q : k_q \neq \hat{k}_q \}$, with $p \triangleq P_r[k_q \neq \hat{k}_q] = \frac{K-1}{K}$. 
	
	Let \(X_q = \mathbb{I}(k_q \neq \hat{k}_q)\) be a Bernoulli random variable with success probability \(p\), representing an error in the \(q\)-th group's index. Thus, \(\zeta^{\prime} = \sum_{q=1}^{N_Q} X_q \sim \text{Binomial}(N_Q, p)\), where \(\zeta^{\prime} \in \{0, 1, \ldots, N_Q\}\). The random variable \(\zeta\) in \eqref{eq:zeta_def}, defined as
	\begin{equation}
		\zeta \triangleq \zeta^{\prime} \mid \zeta^{\prime} > 0,
	\end{equation}
follows a truncated binomial distribution with parameters \(N_Q\) and \(p = (K-1)/K\). Thus, the probability mass function (PMF) of $\zeta$ is given by
	\begin{align}
		\nonumber
		\Pr[\zeta = k] &= \Pr[\zeta' = k \mid k > 0] \\ \nonumber
		& = \frac{\binom{N_Q}{k} p^k (1-p)^{N_Q-k}}{\Pr[k > 0]} = \frac{\binom{N_Q}{k} p^k (1-p)^{N_Q-k}}{1 - (1-p)^{N_Q}} \\
		&= \frac{\binom{N_Q}{k} (K-1)^\zeta}{K^{N_Q}-1} , \quad k = 1, 2, \ldots, N_Q
		\label{eq:pmf}
	\end{align}

	Substituting \eqref{eq:inner_exp2} into \eqref{eq:xxxx} and taking the expectation with respect to $\zeta$ with PMF in \eqref{eq:pmf}, the average PEP becomes:
	\begin{equation}
		\bar{P}_{\mathbf{k} \hat{\mathbf{k}}} = \mathbb{E}_{\zeta} \left[ F(\zeta) \right] = \sum_{\zeta=1}^{N_Q}  \frac{\binom{N_Q}{k} (K-1)^\zeta}{K^{N_Q}-1} F(\zeta)
		\label{eq:final_avg_SER_1}
	\end{equation}
	
Applying the union bound \eqref{eq:unionbound} with $|\mathcal{K}|=K^{N_Q}$, assuming symmetry across all $\bar{P}_{\mathbf{k} \hat{\mathbf{k}}}$, we obtain:
\begin{align}
    P_e &\lesssim \frac{1}{N_Q K^{N_Q}} \sum_{\mathbf{k}\in\mathcal{K}} \sum_{\substack{\hat{\mathbf{k}}\in\mathcal{K}\\\hat{\mathbf{k}}\neq\mathbf{k}}} 
        \bar{P}_{\mathbf{k}\hat{\mathbf{k}}}\,\ell(\mathbf{k},\hat{\mathbf{k}}) 
    \notag\\
    &\stackrel{\text{(a)}}{=} \frac{\bar{P}_{\mathbf{k}\hat{\mathbf{k}}}}{N_Q K^{N_Q}} 
       \sum_{\mathbf{k}\in\mathcal{K}} \sum_{\substack{\hat{\mathbf{k}}\in\mathcal{K}\\\hat{\mathbf{k}}\neq\mathbf{k}}} 
       \ell(\mathbf{k},\hat{\mathbf{k}})
    \notag\\
    &\stackrel{\text{(b)}}{=} \frac{\bar{P}_{\mathbf{k}\hat{\mathbf{k}}}}{N_Q K^{N_Q}} 
       \sum_{\mathbf{k}\in\mathcal{K}} \sum_{\hat{\mathbf{k}}\in\mathcal{K}} 
       \ell(\mathbf{k},\hat{\mathbf{k}})
    \notag\\
    &\stackrel{\text{(c)}}{=} \frac{\bar{P}_{\mathbf{k}\hat{\mathbf{k}}}}{N_Q K^{N_Q}} 
       \sum_{q=1}^{N_Q} \sum_{\mathbf{k}\in\mathcal{K}} \sum_{\hat{\mathbf{k}}\in\mathcal{K}}
       \mathbb{I}(k_q \neq \hat{k}_q)
    \notag\\
    &\stackrel{\text{(d)}}{=} \frac{\bar{P}_{\mathbf{k}\hat{\mathbf{k}}}}{N_Q K^{N_Q}} 
       \sum_{q=1}^{N_Q} \bigl[K^{2N_Q-2}\,K\,(K-1)\bigr]
    \notag\\
    &= \frac{\bar{P}_{\mathbf{k}\hat{\mathbf{k}}}}{N_Q K^{N_Q}}
       \,N_Q\,(K-1)\,K^{2N_Q-1}
    \notag\\
    &= (K-1)\,K^{N_Q-1}\,\bar{P}_{\mathbf{k}\hat{\mathbf{k}}},
    \label{eq:secondpart}
\end{align}
where step (a) follows because the average PEP, \(\bar{P}_{\mathbf{k} \hat{\mathbf{k}}}\), is independent of the indices \(\mathbf{k}\) and \(\hat{\mathbf{k}}\), step (b) follows because \(\ell(\mathbf{k}, \mathbf{k}) = 0\), step (c) uses the definition in equation~\eqref{eq:EuclideanDis}, and step (d) uses combinatorial counting and algebraic simplification as follows. There are \(2N_Q\) total entries across the pair \((\mathbf{k}, \hat{\mathbf{k}})\), since each of \(\mathbf{k}\) and \(\hat{\mathbf{k}}\) is of length \(N_Q\). For a fixed index position \(q\), the remaining \(2N_Q - 2\) entries (i.e., all entries of \(\mathbf{k}\) and \(\hat{\mathbf{k}}\) except \(k_q\) and \(\hat{k}_q\)) can each independently take any of the \(K\) values, resulting in \(K^{2N_Q - 2}\) possible combinations. For the pair \((k_q, \hat{k}_q)\), there are \(K\) choices for \(k_q\), and for each such choice, there are \((K - 1)\) valid choices for \(\hat{k}_q \neq k_q\). Therefore, the total number of index pairs \((\mathbf{k}, \hat{\mathbf{k}})\) satisfying \(k_q \neq \hat{k}_q\) is $  K^{2N_Q - 2} \times K \times (K - 1) = (K - 1)\,K^{2N_Q - 1}$.
This completes the proof.
\end{proof}

Finally, since the average PEPs are independent of the indices $\mathbf{k}$ and $\hat{\mathbf{k}}$, a rough estimation of the BER is given by:
\begin{equation}
	P_b \approx \frac{P_e}{2}
	\label{eq:BER_bound}
\end{equation}

\begin{remark}
	For the special case $N_Q = 1$, i.e., the RIS is treated as a single group, substituting $N_Q=1$ in \eqref{eq:thmbound} we obtain 
	\begin{align}
		P_e &\lesssim (K-1) \left( \frac{1 - \nu}{2} \right)^{n_R} \sum_{k=0}^{n_R-1} \binom{n_R-1+k}{k} \left( \frac{1 + \nu}{2} \right)^k,
		\label{eq:firstpart}
	\end{align}
	where $\nu \triangleq \mu(1) = \sqrt{\frac{NE_s/\sigma^2}{2+NE_s/\sigma^2}}$.
\end{remark}

\section{Asymptotic analysis and capacity evaluation}
\label{sec4}
In this section, we discuss the asymptotic behavior of the SER  derived in the previous section and then derive the capacity. In the capacity analysis, we consider a general case where the transmitter sends an $M$-ary modulation symbol $x \in \mathcal{X}$ (with $|\mathcal{X}| = M$) along with index symbols.

\subsection{Asymptotic error performance and diversity order analysis}

From Proposition~\ref{thm:SymbolErrorBound}, the average PEP bound is expressed as:
\begin{equation}
	P_e \lesssim \frac{K^{N_Q-1}}{K^{N_Q} - 1} \sum_{\zeta=1}^{N_Q} \binom{N_Q}{\zeta} (K-1)^{\zeta+1} F(\zeta),
\end{equation}
where 
\[ F(\zeta) = \left( \frac{1 - \mu(\zeta)}{2} \right)^{n_R} \sum_{k=0}^{n_R-1} \binom{n_R - 1 + k}{k} \left( \frac{1 + \mu(\zeta)}{2} \right)^k, \]
with 
\[ \mu(\zeta) = \sqrt{\frac{N E_s \zeta / \sigma^2}{2 N_Q + N E_s \zeta / \sigma^2}}. \]

Under the high effective SNR condition \( \frac{N}{ N_Q}\cdot \frac{E_s}{\sigma^2} \gg 1 \):
\begin{itemize}
	\item Rewrite \( \mu(\zeta) = \left(1 + \frac{2 N_Q}{\zeta N E_s / \sigma^2}\right)^{-1/2} \). Using the Taylor expansion \( (1 + x)^{-1/2} \approx 1 - \frac{x}{2} \) for small \( x \), we get:
	\[ \mu(\zeta) \approx 1 - \frac{N_Q}{\zeta N E_s / \sigma^2}. \]
	Thus,
	\[ \frac{1 - \mu(\zeta)}{2} \approx \frac{N_Q \sigma^2}{2 \zeta N E_s}, \]
	so
	\[ \left( \frac{1 - \mu(\zeta)}{2} \right)^{n_R} \approx \left( \frac{2 N E_s}{ N_Q \sigma^2} \right)^{-n_R} \zeta^{-n_R}. \]
	\item Since \( \frac{1 + \mu(\zeta)}{2} \approx 1 \), the sum simplifies:
	\[ \sum_{k=0}^{n_R-1} \binom{n_R - 1 + k}{k} \left( \frac{1 + \mu(\zeta)}{2} \right)^k \approx \sum_{k=0}^{n_R-1} \binom{n_R - 1 + k}{k}. \]
	\item By the binomial identity \( \binom{n}{k} = \binom{n}{n-k} \) and the Hockey-Stick Identity~\cite{GrahamKnuthPatashnik1994},
	\[ \sum_{k=0}^{n_R-1} \binom{n_R - 1 + k}{k} = \binom{2 n_R - 1}{n_R - 1} = \frac{1}{2} \binom{2 n_R}{n_R}. \]
\end{itemize}

Thus, 
\[ F(\zeta) \approx \frac{1}{2} \binom{2 n_R}{n_R} \left( \frac{2 N E_s}{ N_Q \sigma^2} \right)^{-n_R} \zeta^{-n_R}, \]
and the error probability bound reduces to:
\begin{align}
    P_e \approx c \left( \frac{2 N E_s}{N_Q \sigma^2} \right)^{-n_R} &=   \left(c^{-\frac{1}{n_R}} \frac{2 N}{N_Q}\right)^{-n_R} \mathrm{SNR}^{-n_R} \nonumber \\
        &={G_c}^{-n_R} \mathrm{SNR}^{-n_R},
    \label{eq:asymptotic1}
\end{align}

where $\mathrm{SNR}=E_s/\sigma^2$ and $c$ is defined by
\begin{equation} 
	c = \frac{(K-1) K^{N_Q-1}}{2 (K^{N_Q} - 1)} \binom{2 n_R}{n_R} \sum_{\zeta=1}^{N_Q} \binom{N_Q}{\zeta} (K-1)^\zeta \zeta^{-n_R}.
\label{eq:def_constant}
\end{equation}

Thus, from \eqref{eq:asymptotic1} we can see that the diversity gain of the system is $n_R$ and the coding gain is expressed as 
\begin{equation}
	G_c =  \frac{2N}{N_Q} c^{-\frac{1}{n_R}}
\end{equation}

For the single-group case (\(N_Q = 1\)), the error probability asymptotically converges to:
\begin{equation}
P_e \approx \frac{(K - 1)}{2 (2 N)^{n_R}} \binom{2 n_R}{n_R}  \mathrm{SNR}^{-n_R}.
	\label{eq:asymptotic2}
\end{equation}

\subsection{Impact of RIS partitioning on error performance}
\label{impact_partition}
For two system configurations with identical SE, we analyze the error probability gap between a system with $N_Q > 1$ (partitioned RIS) and one with $N_Q = 1$ (no partitioning). 

Consider \textit{system 1} with $N_Q > 1$, $K \triangleq K_1$, and hence $R_1 = N_Q \log_2(K_1)$, and \textit{system 2} with $N_Q = 1$, $K \triangleq K_2 = K_1^{N_Q}$, and hence $R_2 = \log_2(K_2) = N_Q \log_2(K_1)$ (i.e., $R_1=R_2$). 

For \textit{systems 1} and \textit{2},  $P_{e1} \approx c_1 \left( \frac{2 N E_s}{N_Q \sigma^2} \right)^{-n_R}$ and $P_{e2} \approx c_2 \left( \frac{2 N E_s}{\sigma^2} \right)^{-n_R}$, respectively, where $c_1 = \frac{(K_1 - 1) K_1^{N_Q - 1}}{2 (K_1^{N_Q} - 1)} \binom{2 n_R}{n_R} \sum_{\zeta=1}^{N_Q} \binom{N_Q}{\zeta} (K_1 - 1)^\zeta \zeta^{-n_R}$ and $c_2 = \frac{1}{2} \binom{2 n_R}{n_R} (K_1^{N_Q} - 1)$. 

The ratio $\frac{P_{e1}}{P_{e2}} = N_Q^{n_R} \frac{c_1}{c_2}$ simplifies to
\begin{align}
\nonumber
	&\Gamma_e(n_R, N_Q, K_1) \triangleq \frac{P_{e1}}{P_{e2}} \nonumber \\ \nonumber
	&= N_Q^{n_R}   \frac{(K_1 - 1) K_1^{N_Q - 1}}{(K_1^{N_Q} - 1)^2} \sum_{\zeta=1}^{N_Q} \binom{N_Q}{\zeta} (K_1 - 1)^\zeta \zeta^{-n_R} \\
& \ge 1,
	\label{eq:errorgap}
\end{align}
indicating that, at the same SE, RIS partitioning results in a higher error probability compared with the no-partitioning case. This means that, although both systems exhibit the same BER decay rate, there remains a fixed asymptotic gap between their BER curves.

\subsection{Capacity analysis}

Rewrite the received signal:
\begin{equation}
	\mathbf{y} = \sqrt{E_s} \mathbf{d}_{\mathbf{k}} x + \mathbf{n}
\end{equation}
where $\mathbf{d}_{\mathbf{k}} = \sum_{q=1}^{N_Q} \mathbf{G}_q \operatorname{diag}(\mathbf{h}_q) \mathbf{s}_{k_q}$ as defined in \eqref{eq:effective_channel_GCSM}, $\mathbf{k}= [k_1,k_2,\cdots, k_{N_Q}]^T \in \mathcal{K}$ (with $|\mathcal{K}|=K^{N_Q}$) is the index vector and $x \in \mathcal{X}$ is $M$-ary modulation symbol (with normalized power) sent by the transmitter alongside index information.

Conditioned on the subchannels $\mathbf{G}_q,\mathbf{h}_q$($q=1,\cdots, N_Q$), the mutual information between the input pair \((\mathbf{k}, x)\) and the output $\mathbf{y}$ is given by:
\begin{equation}
	I(\mathbf{k}, x; \mathbf{y}) = \mathbb{E} \left[ \log_2 \frac{p(\mathbf{y} \mid \mathbf{k}, x)}{p(\mathbf{y}) } \right],
	\label{eq:mutualinfo}
\end{equation}
where the expectation is over the joint distribution $p(\mathbf{k}, x, \mathbf{y})$. Assuming independent and uniform inputs for the index vector $\mathbf{k}$ and modulation symbol $x$, we have $p(\mathbf{k}) = \frac{1}{K^{N_Q}}$ and $p(x) = \frac{1}{M}$, leading to the joint probability 
\begin{equation}
	p(\mathbf{k}, x) = \frac{1}{K^{N_Q} M}.
\end{equation}

The conditional probability distribution of the received signal given $\mathbf{k}$ and $x$ is:
\begin{equation}
	p(\mathbf{y} \mid \mathbf{k}, x) = \frac{1}{(\pi \sigma^2)^{n_r}} \exp\left( -\frac{1}{\sigma^2} \|\mathbf{y} - \sqrt{E_s} \mathbf{d}_{\mathbf{k}} x\|^2 \right).
	\label{eq:condi_prob_cap}
\end{equation}
The marginal probability distribution of the received signal is obtained by averaging over all possible input pairs $(\mathbf{k},x) \in \mathcal{K} \times \mathcal{X}$:
\begin{align}
	\nonumber
	&p(\mathbf{y}) = \sum_{\mathbf{k}\in \mathcal{K}} \sum_{x \in \mathcal{X}} p(\mathbf{k}, x) p(\mathbf{y} \mid \mathbf{k}, x) \\
	&= \frac{1}{K^{N_Q} M} \sum_{\mathbf{k}} \sum_{x} \frac{1}{(\pi \sigma^2)^{n_r}} \exp\left( -\frac{1}{\sigma^2} \|\mathbf{y} - \sqrt{E_s} \mathbf{d}_{\mathbf{k}} x\|^2 \right).
	\label{eq:marginal_prob}
\end{align}

Substituting \eqref{eq:condi_prob_cap} and \eqref{eq:marginal_prob} into \eqref{eq:mutualinfo}, the mutual information can be expressed as
\begin{equation}
	\begin{split}
		&	I(\mathbf{k}, x; \mathbf{y})= \log_2 (M K^{N_Q})-  \\
		&  \frac{1}{K^{N_Q} M (\pi \sigma^2)^{n_r}} \sum_{\mathbf{k} \in \mathcal{K}} \sum_{x \in \mathcal{X}}  \int  \exp\left( -\frac{1}{\sigma^2} \|\mathbf{y} - \sqrt{E_s} \mathbf{d}_{\mathbf{k}} x\|^2 \right) \times \\
		& \log_2 \left [ \frac{ \sum_{\mathbf{k}^{\prime} \in \mathcal{K}} \sum_{x^{\prime \in \mathcal{X}}} \exp\left( -\frac{1}{\sigma^2} \|\mathbf{y} - \sqrt{E_s} \mathbf{d}_{\mathbf{k}^{\prime}} x^{\prime}\|^2 \right)}{ \exp\left( -\frac{1}{\sigma^2} \|\mathbf{y} - \sqrt{E_s} \mathbf{d}_{\mathbf{k}} x\|^2 \right)  }\right] \, d\mathbf{y}.
	\end{split}
	\label{eq:mutual_information_split_equation}
\end{equation}

The mutual information \eqref{eq:mutual_information_split_equation} is bounded as
$I(\mathbf{k}, x; \mathbf{y}) \leq N_Q \log_2 K + \log_2 M$, with equality achieved at sufficiently high SNR as the second term in \eqref{eq:mutual_information_split_equation} approaches 0. The system capacity is defined as the expectation of  $I(\mathbf{k}, x; \mathbf{y})$, averaged over the channel realizations $\{\mathbf{G}_q,  \mathbf{h}_q\}, q=1,2,\cdots,N_Q$:
\begin{equation}
	C = \mathbb{E}\left[ I(\mathbf{k}, x; \mathbf{y}) \right],
	\label{eq:syscapacity}
\end{equation}

For the case of unmodulated carrier, we set $M=1$ and $x=1$. Given the analytical intractability, the final expression \eqref{eq:syscapacity} is usually found numerically (for small $|\mathcal{K}| |\mathcal{X}|$) or through Monte Carlo simulation.

\section{Simulation Results}
\label{sec5}

In this section, we assess the performance of the proposed RIS-CSM scheme and validate the analytical results under various configurations. We consider a dual-hop RIS-assisted system with a single-antenna transmitter, a multi-antenna receiver, and a blocked direct Tx–Rx link, where the RIS mediates the transmission. Both the Tx–RIS and RIS–Rx channels are modeled as flat Rayleigh fading with IID $\mathcal{CN}(0,1)$ channel gains, constant over one transmission and independently varying between transmissions. Unless stated otherwise, RIS elements are assumed uncorrelated. For all scenarios, the RIS phase-shift vectors are generated from a Hadamard matrix with appropriate dimensions, as described in Sec.~\ref{sec2}. The receiver is assumed to have perfect knowledge of the effective channels in \eqref{eq:rx_signal}, and, unless otherwise specified, the transmitter in RIS-CSM sends an unmodulated carrier.

\subsection{SER performance with varying RIS size}
\begin{figure}[t]
	\centering
	\includegraphics[width=1\columnwidth]{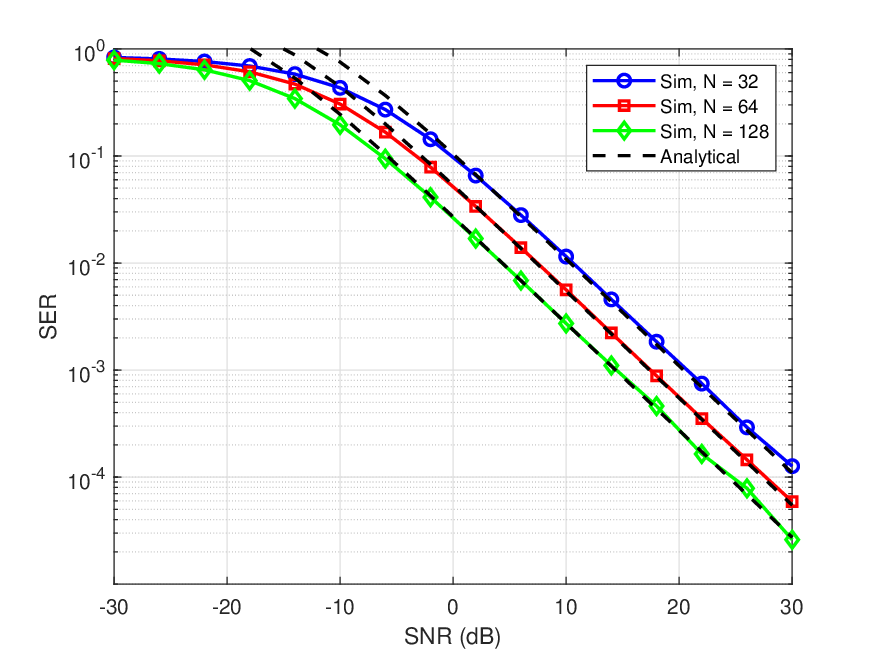}
\caption{SER performance of RIS-CSM at spectral efficiency of 3 bpcu with $n_R=1$, $N_Q=1$, and $K=8$. Analytical results are based on the upper bound in~\eqref{eq:thmbound}.}
	\label{fig:fig3}
\end{figure}

Fig.~\ref{fig:fig3} shows the SER of the RIS-CSM system with a single-antenna receiver ($n_R = 1$). We vary the number of RIS elements ($N = 32, 64, 128$) while keeping the number of phase-shift configurations fixed at $K = 8$, yielding a data rate of $R = \log_2(8) = 3$ bpcu. As expected, the SER clearly improves as $N$ increases. The analytical approximate union bound from \eqref{eq:thmbound} closely matches the simulation results across most SNR values, diverging at very low SNR due to overlapping error events not captured by the union bound. From the asymptotic analysis, the system's diversity is 1, as shown in Fig.~\ref{fig:fig3}.

\subsection{BER Improvement through receiver diversity}

\begin{figure}[t]
	\centering 
	\includegraphics[width=1\columnwidth]{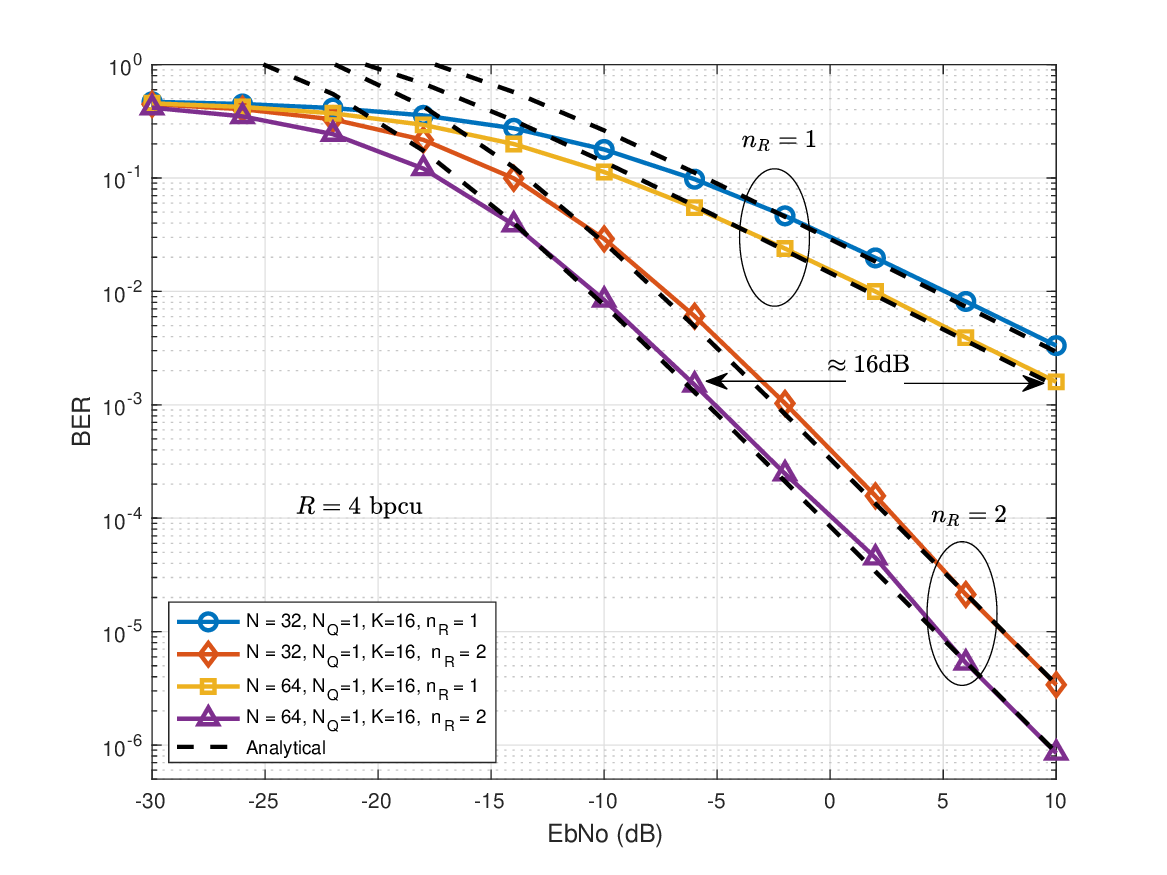}
	\caption{BER performance of RIS-CSM at spectral efficiency of 4 bpcu with $N=32, 64$ and $n_R=1, 2$. Analytical results are derived from \eqref{eq:BER_bound}.}

	\label{fig:fig4}
\end{figure}

Figure~\ref{fig:fig4} shows the bit error rate (BER) performance of the RIS-CSM system versus the bit-energy-to-noise power spectral density ratio ({EbNo}),  where \(N_Q = 1\) and \(K = 16\), yielding a SE of 4 bpcu. The analytical results are obtained from \eqref{eq:BER_bound}. The results show a consistent trend of improving BER as the number of RIS elements \(N\) and receive antennas \(n_R\) increase. Notably, adding a second receive antenna significantly enhances performance due to the increase in diversity order from 1 to 2. For example, in a system with \(N = 64\) RIS elements and a target BER of \(10^{-3}\), using two receive antennas instead of one results in a power gain of approximately 16 dB.

\subsection{Impact of RIS partitioning on performance}
\begin{figure}[t]
	\centering 
	\includegraphics[width=1\columnwidth]{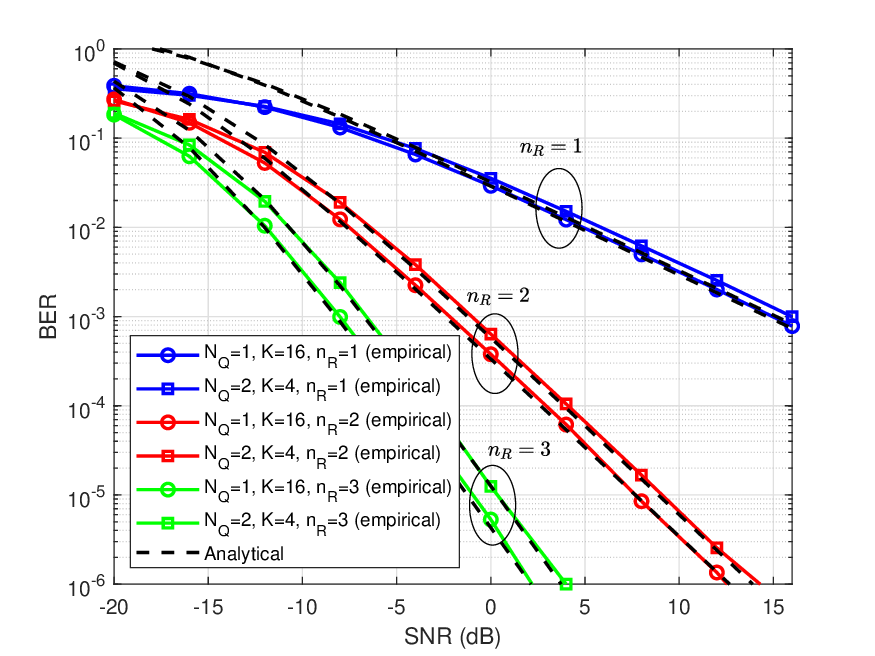}
\caption{Impact of partitioning on the error probability performance of RIS-CSM with $N=128$ and spectral efficiency of 4 bpcu. Analytical results are based on the upper bound in~\eqref{eq:thmbound}.}
	\label{fig:fig5}
\end{figure}

Fig.~\ref{fig:fig5} illustrates the impact of partitioning (grouping) of RIS in RIS-CSM on the BER performance under a SE of 4 bpcu. Two configurations are shown: one with partitioning (\(N_Q = 2, K = 4\)) and one without (\(N_Q = 1, K = 16\)), both designed to achieve the same SE of 4 bpcu. The system without partitioning generally exhibits lower BER, and as the SNR increases, the performance gap becomes almost fixed. This behavior is more pronounced when using 2 or 3 receive antennas. Despite achieving the same SE, this performance gap arises primarily due to interference between groups in the partitioned system and the reduction in the number of elements per group, which affects the power gain. These results are consistent with the asymptotic analysis in Sec.~\ref{sec5}, as seen in \eqref{eq:errorgap}.

As shown in Fig.~\ref{fig:fig5}, at high SNR, the performance gap between RIS configurations with and without partitioning remains nearly constant. This aligns with the theoretical result in \eqref{eq:asymptotic1}, where the asymptotic error probability is given by \( P_e = G_c^{-n_R} \mathrm{SNR}^{-n_R} \), with the constant \( G_c \), defined in \eqref{eq:def_constant}, representing the coding gain, which is independent of SNR. Both systems (with and without partitioning) share the same diversity order \( n_R \), ensuring identical error probability decay rates, as depicted in Fig.~\ref{fig:fig5}. However, the coding gain \( G_c \) differs between configurations, resulting in a constant, SNR-independent error rate gap in the asymptotic regime. As derived in \eqref{eq:errorgap}, the error probability ratio shows that the system with partitioning (\( N_Q = 1 \)) has a higher error rate than with \( N_Q > 1 \), with a fixed BER gap due to differing coding gains.

\subsection{Performance comparison}

Here, the performance of the proposed RIS-CSM scheme is compared against three existing schemes from the literature: RIS-MIMO, RIS-GSM, and RIS-CIM. 
For a fair comparison, all schemes employ a single-antenna transmitter, the same number of receive antennas($n_R=1,2$), identical spectral efficiency of $R=4$ bpcu, and an ML detector, with perfect CSI assumed at the receiver. Moreover, none of the schemes perform RIS-side beamforming, ensuring consistency with the proposed RIS-CSM. Table~\ref{tab:RIS_schemes} summarizes the compared schemes, describing their encoding operations and the corresponding simulation settings for clarity.

\begin{table*}[!t]
\centering
\caption{Comparison of RIS-based modulation schemes (all at $R=4$~bpcu).}
\label{tab:RIS_schemes}
\renewcommand{\arraystretch}{1.3}
\setlength{\tabcolsep}{4pt}
\begin{tabular}{p{2.2cm} p{11.5cm}}
\hline
\multicolumn{1}{c}{\textbf{Scheme}} & \multicolumn{1}{c}{\textbf{Encoding process}} \\
\hline
{RIS-CSM (proposed)} 
& $N_Q=1$; RIS uses $K=16$ phase-shift patterns from the first $16$ columns of a $64\times 64$ Hadamard matrix. Transmitter sends unmodulated carrier; RIS selects pattern to send $R=\log_2(K)=4$~bpcu. No RIS-side beamforming. \\

{RIS-MIMO}~\cite{9535453,9729740} 
& $N_Q=1$; All RIS elements reflect signals with a common phase from a 16-PSK set $e^{j2\pi(m-1)/16}$. Transmitter sends unmodulated carrier; $R=\log_2(16)=4$~bpcu. No RIS-side beamforming. \\

{RIS-GSM}~\cite{9217944,9535453} 
& $N_Q=4$, $N_A=3$ active groups; transmitter sends 4-QAM symbol, only active groups reflect it; $R=2+\lfloor \log_2\binom{4}{3} \rfloor =4$~bpcu. No RIS-side beamforming. \\

{RIS-CIM}~\cite{10138929} 
& $N_Q=1$; transmitter sends 128-QAM symbol, RIS applies one of $W=2$ spreading codes of length $2$ to all elements. $R=\frac{7+1}{2}=4$~bpcu. No RIS-side beamforming. \\
\hline
\end{tabular}
\end{table*}

For clarity, and due to the lack of consensus on terminology, we note that the terms \emph{RIS-MIMO} and \emph{RIS with reflection phase modulation} (RIS-RM) are used interchangeably in the literature to describe the same scheme.
Similarly, \emph{RIS-GSM} is equivalent to \emph{RIS with reflection pattern modulation} (RIS-RPM).
In this work, we adopt the notation RIS-MIMO and RIS-GSM for brevity and to avoid confusion, with the alternative names noted above.

\begin{figure}[t]
	\centering 
	\includegraphics[width=0.92\columnwidth]{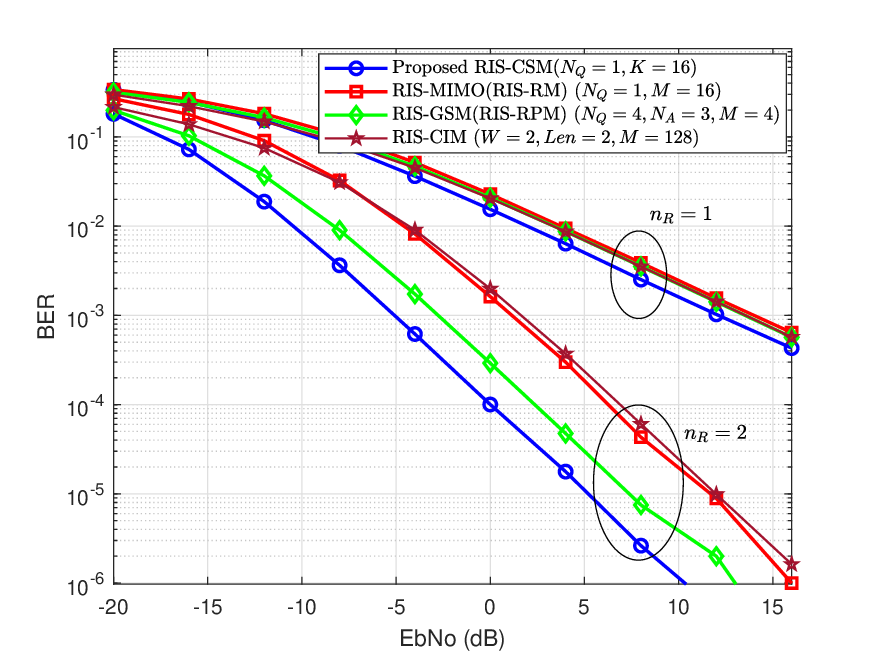}
\caption{BER performance comparison of RIS-CSM, RIS-MIMO (RIS-RM), RIS-GSM (RIS-RPM), and RIS-CIM at a spectral efficiency of 4~bpcu with $N = 64$ and $n_R \in \{1,2\}$. 
RIS-CSM: $N_Q = 1$ partition with $K = 16$ phase-shift patterns; 
RIS-MIMO: 16-PSK modulation at the RIS with a single partition; 
RIS-GSM: $N_Q = 4$ groups with $N_A = 3$ active, reflecting 4-QAM symbols; 
RIS-CIM: $M = 128$-ary modulation with $W = 2$ spreading codes of length~2.}

	\label{fig:fig6}
\end{figure}

Since all schemes operate without RIS-side beamforming under identical settings, their differences mainly arise from robustness and efficiency.

\textit{RIS-MIMO:} Achieves power gain scaling linearly with the total RIS elements $N$, with spectral efficiency $R = \log_2(M_{tx}) + N_Q \log_2(M_{ris})$ (bpcu), where $M_{tx}$ and $M_{ris}$ are the modulation orders at the transmitter and RIS, respectively. However, higher modulation orders make ML detection harder due to closely spaced constellation points and require finer phase resolution at the RIS, complicating implementation.

\textit{RIS-GSM:} Improves spectral efficiency by activating subsets of RIS groups but suffers reduced effective reflecting area from deactivated subsurfaces, lowering received signal strength. The maximum index bits are limited by $\left\lfloor \log_2 \binom{N_Q}{N_Q/2} \right\rfloor$, so spectral efficiency is $R = \log_2(M_{tx}) + \left\lfloor \log_2 \binom{N_Q}{N_Q/2} \right\rfloor$ (bpcu). Large $N_Q$ is often impractical due to control complexity and power loss, reducing spectral efficiency.

\textit{RIS-CIM:} Incurs bandwidth overhead due to spreading codes, reducing spectral efficiency by factor $Len$, yielding $R = \frac{\log_2(M_{tx}) + \log_2(K)}{Len}$ (bpcu), where $W$ is the number of spreading codes. Hence, RIS index bits contribute little, with most spectral efficiency coming from transmitter modulation.

For reference, Table \ref{tab:SE_formulas} shows the formulas for computing the achievable SE of each scheme.
 
\begin{table}[htbp]
\centering
\caption{SE comparison of simulated schemes}
\label{tab:SE_formulas}
\renewcommand{\arraystretch}{1.3}
\begin{tabular}{lc}
\hline
\textbf{Scheme} & \textbf{SE (bpcu)} \\
\hline
RIS-MIMO & $R = \log_2(M_{tx}) + N_Q \log_2(M_{ris})$ \\

RIS-GSM & $R = \log_2(M_{tx}) + \left\lfloor \log_2 \binom{N_Q}{N_A} \right\rfloor$ \\

RIS-CIM & $R = \frac{\log_2(M_{tx}) + \log_2(W)}{Len}$ \\

Proposed RIS-CSM & $R = N_Q \log_2(K) + \log_2(M_{tx})$ \\
\hline
\end{tabular}
\end{table}

As shown in Fig.~\ref{fig:fig6}, the BER performance of all schemes improves significantly when increasing the number of receive antennas from $n_R = 1$ to $n_R = 2$, which is expected. 
For $n_R = 1$, RIS-MIMO (RIS-RM), RIS-GSM (RIS-RPM), and RIS-CIM exhibit nearly identical performance. RIS-MIMO and RIS-GSM maintain similar performance in the single-antenna case, but their performance gap becomes more pronounced when transitioning to two receive antennas as diversity increases for all schemes, 
whereas RIS-CIM performs almost the same as RIS-MIMO in both cases. 
In all considered scenarios, RIS-CSM consistently achieves better BER performance than other schemes. This is attributed to larger distances between channel signature constellation points, resulting in improved distinguishability by the ML detector.

\subsection{Effect of imperfect CSI on performance}
\begin{figure}[t]
	\centering 
	\includegraphics[width=0.92\columnwidth]{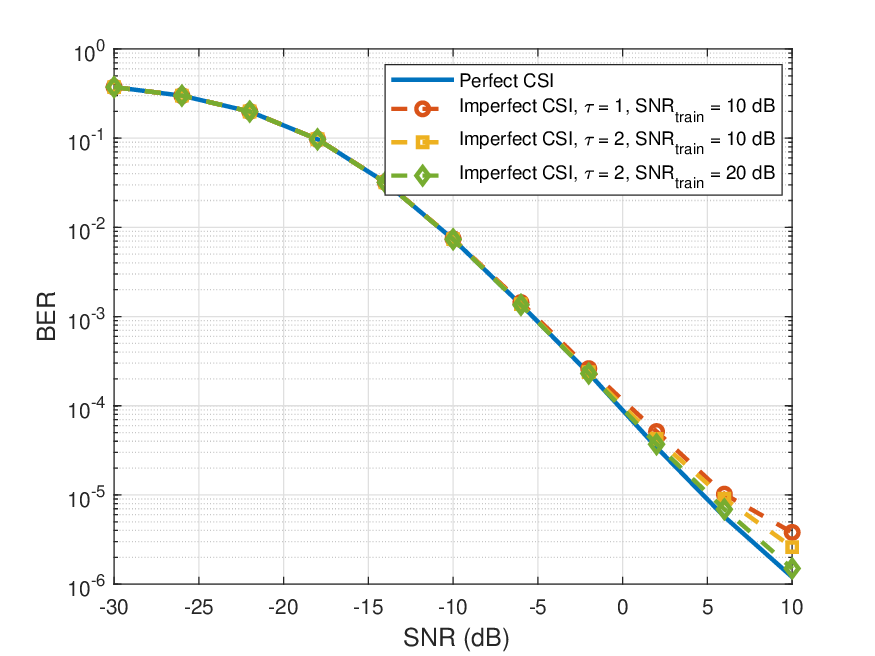}
\caption{BER performance of a 4-bit RIS-CSM system with MMSE channel estimation, for $N=128$, $K=16$, and $n_R=2$.}
	\label{fig:fig7}
\end{figure}

Fig.~\ref{fig:fig7} shows the BER of a 4-bit RIS-CSM system under perfect and imperfect CSI. For channel estimation, the MMSE estimator is used, as described in Sec.~\ref{sec2}, where the effective channels incorporating phase-shift vectors are estimated. For $\tau=1$ and $\tau=2$, the receiver collects $K=16$ and $2K=32$ measurements, respectively, each with a dimension of 2. As shown, the BER improves with increasing $\tau$ and training signal-to-noise ratio. At simulated SNR = 15\,dB with $\tau=2$, the BER is nearly identical to that of the perfect CSI case across the considered SNR range.

\begin{figure}[t]
	\centering
	\includegraphics[width=1\columnwidth]{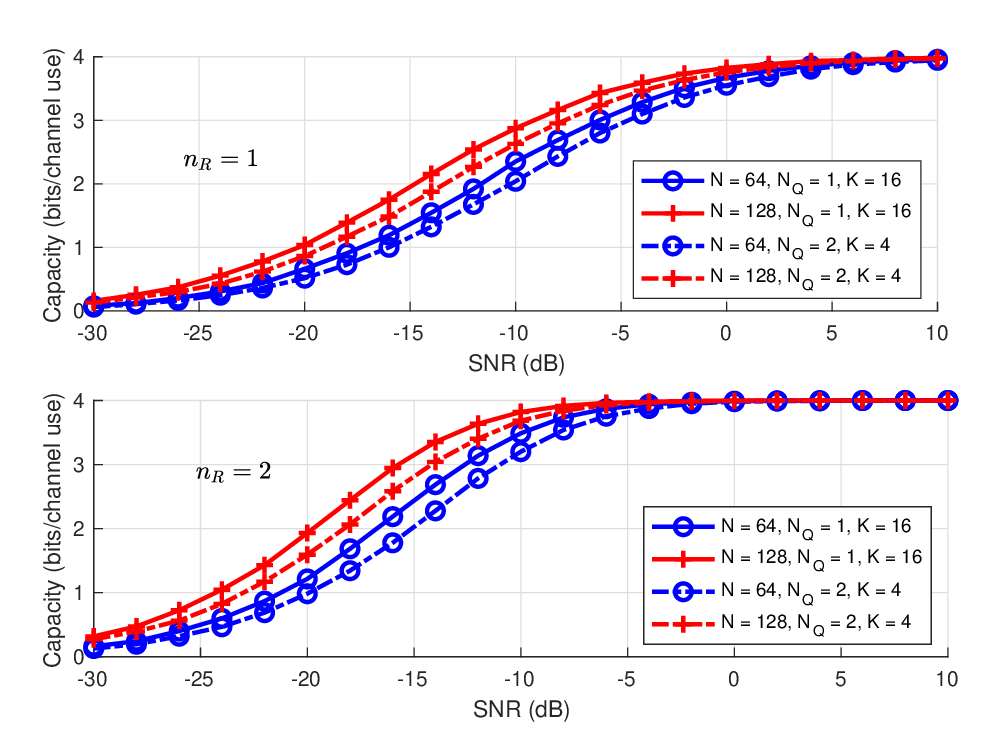}
\caption{Capacity comparison of the RIS-CSM system with an unmodulated carrier at the transmitter, operating at a maximum spectral efficiency of 4 bpcu, with and without RIS partitioning: (top) \(n_R=1\) (single receive antenna); (bottom) \(n_R=2\) (two receive antennas).}

	\label{fig:fig8}
\end{figure}

\subsection{Capacity evaluation}
\begin{figure}[t]
	\centering
	\includegraphics[width=1\columnwidth]{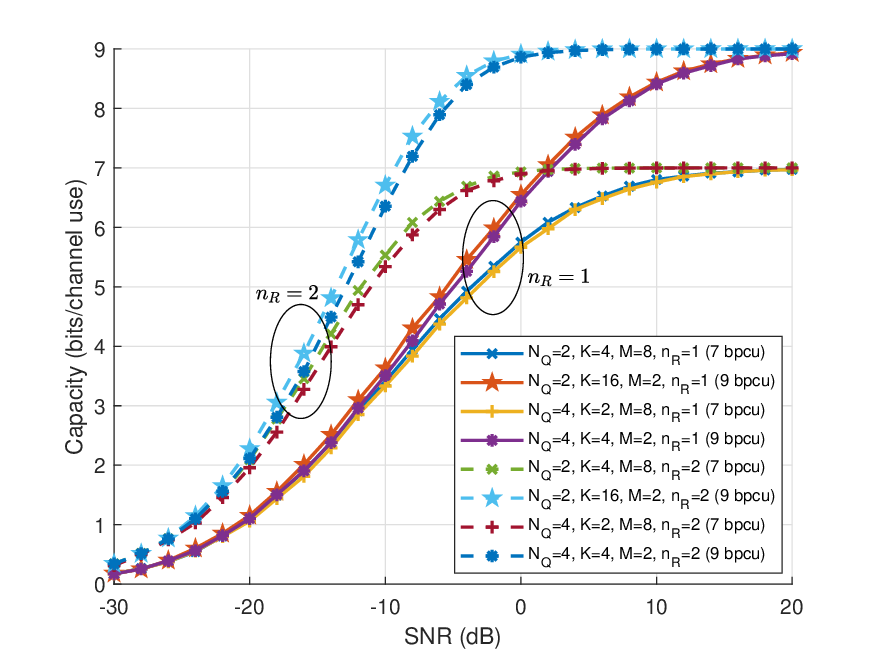}
\caption{Capacity of the RIS-CSM system with $M$-ary modulation at the transmitter, shown for two RIS partitioning configurations ($N_Q=2$ and $N_Q=4$) at high spectral efficiency levels. The total number of RIS elements is $N=128$.}
	\label{fig:fig9}
\end{figure}

Figs.~\ref{fig:fig8} and~\ref{fig:fig9} depict capacity results for the RIS-CSM system. In 	Fig.~\ref{fig:fig8}, we consider comparing configurations with and without partitioning under a maximum SE of 4~bpcu. The figure explores the impact of varying the number of RIS elements (\(N = 64, 128\)) and receive antennas (\(n_R = 1, 2\)). As anticipated, capacity increases with larger \(N\) and \(n_R\) across all configurations. For instance, in the configuration without partitioning (\(N_Q = 1, N = 128\)), the capacity at $\mathrm{SNR}=-10$ dB increases from approximately 2.63~bpcu to 3.82~bpcu when adding a second receive antenna, a gain consistent with the intuitive benefits of increased system diversity. Comparing the upper (\(n_R = 1\)) and lower (\(n_R = 2\)) plots reveals that the configuration without partitioning consistently achieves a higher capacity than the partitioned setup. This capacity gap diminishes as SNR increases. This performance difference arises from the error probability gap between the two system types as discussed under Fig.~\ref{fig:fig5} and as analytically shown under Sec. \ref{sec4}. Consequently, the partitioned system (e.g., with \(N_Q = 2\)) requires extra power to reach the same capacity as the non-partitioned system, demonstrating the latter's greater power utilization in the low-to-mid SNR range. Ultimately, both configurations converge towards the maximum capacity of 4~bpcu at higher SNRs.

In Fig.~\ref{fig:fig9}, detailed capacity comparisons for the RIS-CSM system with $N=128$ RIS elements and partitioning are presented, focusing on high SE levels of 7 and 9 bpcu. The transmission scheme employs \(M\)-ary modulation symbols concurrently with index symbols. To focus on the impact of the RIS partitioning on the capacity comparison, the modulation order \(M\) is held constant for each fixed SE level, ensuring that observed differences are primarily due to the number of partitions (\(N_Q\)). Consistent with prior observations (e.g., from Fig.~\ref{fig:fig8}), an increase in capacity is evident when employing two receive antennas compared to a single antenna. Furthermore, for a given SE, a discernible difference in capacity is observed between systems utilizing \(N_Q = 2\) and \(N_Q = 4\) partitions. While this behavior aligns with the trend noted in Fig.~\ref{fig:fig8}, the magnitude of this capacity disparity is less pronounced at these elevated SE values. From an information-theoretic standpoint, the results in Figs.~\ref{fig:fig8} and~\ref{fig:fig9} suggest that for a fixed SE, a configuration with a smaller number of partitions generally achieves a higher channel capacity.

 \subsection{Impact of spatial correlation on performance}
\begin{figure*}[htbp]
    \centering
    \includegraphics[width=\textwidth]{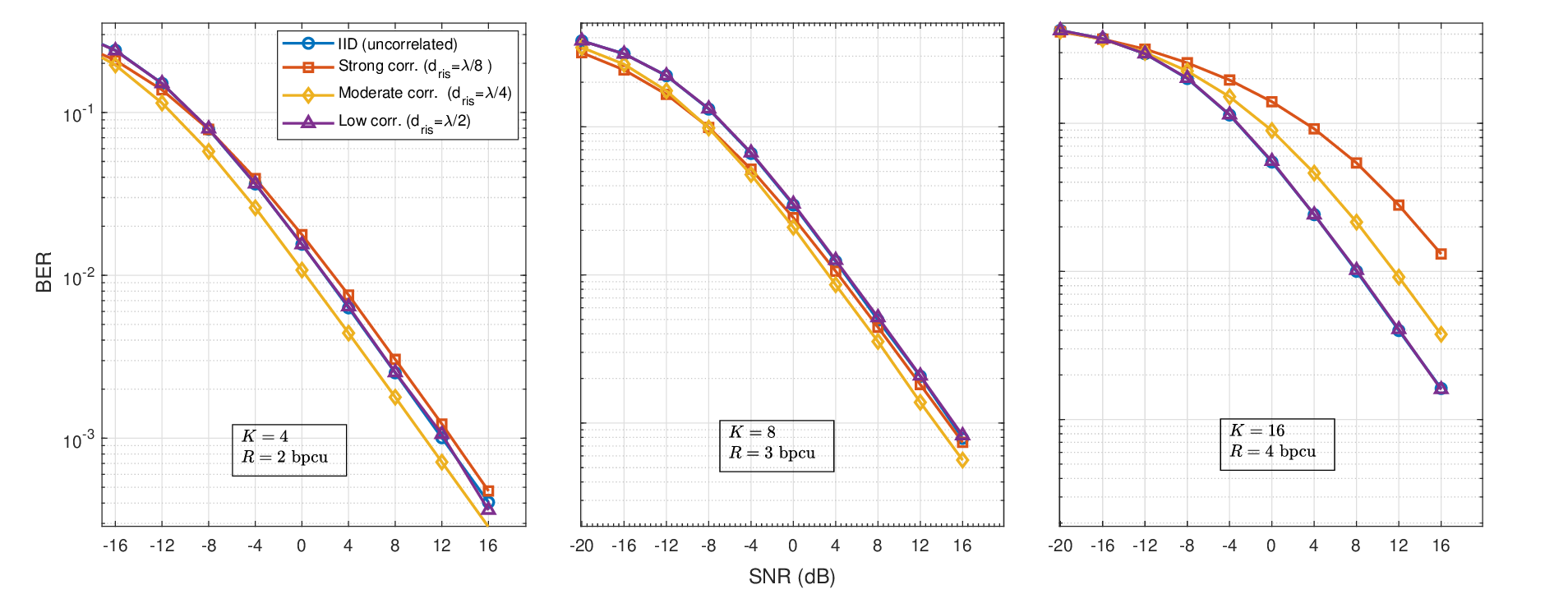}  
\caption{Impact of RIS spatial correlation on system performance for varying element spacings and data rates, with $N=64$, $N_Q=1$, and $n_R=1$.}
    \label{fig:fig10}
\end{figure*}

\begin{figure}[t]
	\centering
	\includegraphics[width=1\columnwidth]{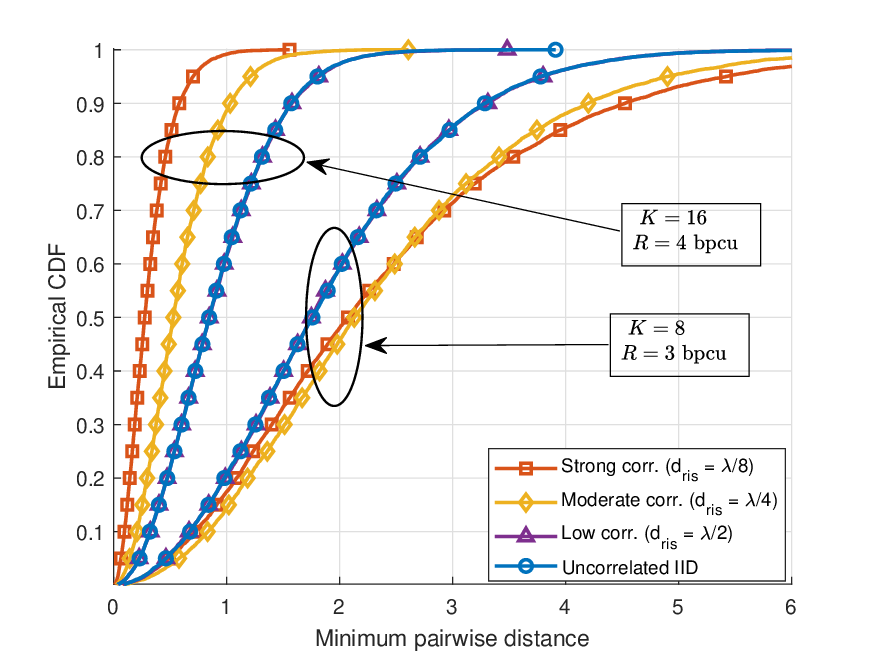}  
\caption{CDF of the minimum pairwise distance of channel signature constellation under spatial correlation for different RIS element spacings and numbers of phase-shift patterns, with $N=64$, $N_Q=1$, and $n_R=1$. The uncorrelated IID case is also included for comparison.}
     \label{fig:fig11}
\end{figure}

\begin{figure}[t]
	\centering
	\includegraphics[width=1\columnwidth]{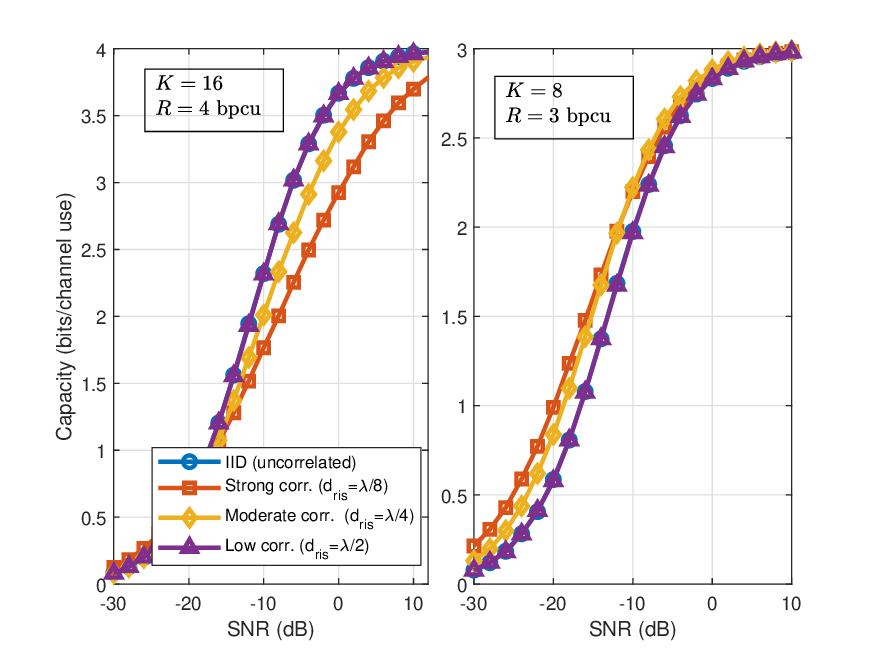}  
\caption{System capacity under spatial correlation for different element spacings and data rates, with $N = 64$, $N_Q = 1$, and $n_R = 1$. 
The left plot corresponds to $K = 16$ (4~bpcu) and the right plot to $K = 8$ (3~bpcu).}
    \label{fig:fig12}
\end{figure}

To gain insight into the impact of spatial correlation on system performance, Figs.~\ref{fig:fig10}-~\ref{fig:fig12} illustrate its effect on the BER and capacity, respectively. We consider a square RIS with $N = 64$ elements arranged in an $8 \times 8$ grid ($N_h = N_v = 8$), with element spacing $d_\text{ris} \in \{\lambda/8, \lambda/4, \lambda/2\}$, where $\lambda$ is the carrier wavelength. Spatial correlation is modeled by the matrix $\mathbf{R}$ with its $(m,n)$-th entry (correlation coefficient) given by \cite{Bjornson21}
\begin{equation}
{r}_{m,n} = \text{sinc}\left(\frac{2 d_\text{ris}} {\lambda} d_{m,n}\right),
\end{equation}
where $d_{m,n}$ is the normalized distance between elements $m$ and $n$, i.e., $d_{m,n} = 1$ for adjacent elements.

\begin{table}[h]
\centering
\caption{Simulation parameters for spatial correlation in Figs.~\ref{fig:fig10} and \ref{fig:fig11}.}
\begin{tabular}{|l|l|}
\hline
\textbf{Parameter} & \textbf{Description} \\
\hline
RIS size ($N_h \times N_v$) & $8 \times 8$ ($N = 64$) \\
Rx/Tx antennas ($n_R$, $n_t$) & 1,\ 1 \\
Phase patterns ($K$) & 4,\ 8,\ 16 \\
Element spacing ($d_\text{ris}$) & $\lambda/8,\ \lambda/4,\ \lambda/2$ \\
Element coordinates ($\mathbf{u}_l$) & using 2D grid \\
Normalized distance ($d_{m,n}$) & $\lVert \mathbf{u}_m - \mathbf{u}_n \rVert$ \\
Correlation coefficient ($r_{m,n}$) & $\text{sinc}\left(\frac{2 d_\text{ris}} {\lambda} d_{m,n}\right)$ \\
Tx-RIS correlated channel ($\tilde{\mathbf{h}}$) & $\mathbf{R}^{1/2} \mathbf{h},\ \mathbf{h} \in \mathbb{C}^{N \times 1}$ \\
RIS-Rx correlated channel ($\tilde{\mathbf{G}}$) & $\mathbf{G} \mathbf{R}^{1/2},\ \mathbf{G} \in \mathbb{C}^{n_R \times N}$ \\
\hline
\end{tabular}
\label{tab:parameters_fig10}
\end{table}

We classify the values $d_\text{ris} \in \{\lambda/8,\ \lambda/4,\ \lambda/2\}$ as corresponding to strong, moderate, and low correlation, respectively. For instance, when $d_\text{ris} = \lambda/8$, the RIS elements are densely packed, with adjacent-element correlation $\mathrm{sinc}(0.25) \approx 0.97$, indicating strong correlation. The correlated Tx-RIS and RIS-Rx channels are generated as $\tilde{\mathbf{h}} = \mathbf{R}^{1/2} \mathbf{h}$ and $\tilde{\mathbf{G}} = \mathbf{G} \mathbf{R}^{1/2}$, where $\mathbf{h} \in \mathbb{C}^{N \times 1}$ and $\mathbf{G} \in \mathbb{C}^{n_R \times N}$ are IID Rayleigh fading channels. Simulation parameters are summarized in Table~\ref{tab:parameters_fig10}.

As shown in Fig.~\ref{fig:fig10}, spatial correlation exhibits both beneficial and detrimental effects on RIS-CSM performance. For small $K$ values (e.g., $K = 4, 8$, corresponding to low spectral efficiency), moderate correlation (e.g., $d_{\text{ris}} = \lambda/4$) outperforms the IID case ($\mathbf{R} = \mathbf{I}_N$) by enhancing the received SNR through constructive alignment of correlated channel coefficients, thus improving ML detection accuracy. This occurs because, for small $K$, the channel signatures which are random(i.e., effective constellation points) are, on average, relatively well-separated, resulting in a sufficiently large expected minimum Euclidean distance. Spatial correlation in this case reduces randomness and variance in the channel, yielding a more structured and predictable constellation that aids detection. However, as $K$ increases (e.g., $K = 16$, higher spectral efficiency), the signal space becomes crowded. In this regime, spatial correlation reduces channel diversity and hence the loss of diversity outweighs the SNR gain, as the receiver cannot reliably distinguish between the many possible signals. 

To support the above result, Fig.~\ref{fig:fig11} shows the cumulative distribution function (CDF) of the minimum Euclidean distance of the channel signature constellation for \(K=8\) and \(K=16\). For \(K=8\), correlation yields larger minimum pairwise distances than the IID case, where points are closer—for instance, under moderate correlation, the probability of distances below 3 is about 72\%, compared to 85\% for IID. However, for \(K=16\), the IID case outperforms the correlated scenario with generally larger minimum distances despite tighter constellation packing. These results highlight a trade-off introduced by spatial correlation between SNR gain and channel diversity, dependent on system parameters.

In Fig.~\ref{fig:fig12}, the system capacity is presented for various spatial correlation levels alongside the IID case. For the case $K=8$ and $R=3$~bpcu (rightmost plot), it can be observed that spatial correlation can enhance capacity: under low SNR, the low-correlation scenario outperforms the moderate-correlation and IID cases, where moderate-correlation case subsequently dominates at higher SNR, while the IID case yields the lowest capacity. When increasing to $K$ and $R=4$~bpcu, low correlation offers a marginal advantage in extremely noisy conditions; however, the IID case provides higher capacity overall compared to the correlated-channel scenarios. This trend is consistent with the BER results in Fig.~\ref{fig:fig10}, where correlation tends to reduce BER for small $K$, whereas for larger $K$, the IID case achieves lower BER. This correspondence between error rate and capacity aligns with information-theoretic principles.

A full characterization of spatial correlation effects is beyond the scope of this work, as it depends on system architecture and modulation in a complex way. However, similar trends are observed in \cite{Guo2023}, where correlation improves performance in RIS systems without phase optimization, as in our case.

\section{Conclusion}
\label{sec6}

This paper proposed RIS-CSM, a simple index modulation scheme for RIS-assisted communication systems with a single-RF transmitter and a receiver equipped with \(n_R\) antennas. By partitioning RIS elements into groups and leveraging orthogonal phase-shift vectors from a Hadamard matrix—where each element's reflection phase is either \(0\) or \(\pi\)—RIS-CSM generates distinct channel signatures to embed information into their indices without requiring RIS-side beamforming or CSI at the RIS. A low-complexity channel estimation method was presented, alongside a closed-form union bound on the error probability, validated through simulations. The system achieves a diversity gain of \(n_R\) and a coding gain that scales with the number of RIS elements. The results demonstrate RIS-CSM’s potential for energy-efficient and cost-effective wireless networks. Moreover, this work highlights the impact of spatial correlation among RIS elements on performance, revealing a trade-off between SNR enhancement and channel diversity, where correlation can enhance performance under low spectral efficiency. Future research could investigate simple strategies to incorporate limited beamforming with minimal CSI at the RIS and explore integration with other index modulation techniques to further improve performance.

\bibliographystyle{elsarticle-num}
\bibliography{references}
\end{document}